\documentclass[prb,twocolumn,english,aps,superbib,tightenlines,floatfix,superscriptaddress]{revtex4-1}

\usepackage{graphicx,epsfig}
\usepackage{hyperref}

\begin{document}

\newcommand{\half}{\frac{1}{2}}
\title{Lagrange formalism of memory circuit elements: Classical and quantum formulations}
\author{Guy Z. Cohen}
\email{gcohen@physics.ucsd.edu}
\affiliation{Department of
Physics, University of California, San Diego, La Jolla, California
92093-0319}

\author{Yuriy V. Pershin}
\email{pershin@physics.sc.edu}
\affiliation{Department of Physics
and Astronomy and USC Nanocenter, University of South Carolina,
Columbia, SC, 29208}

\author{Massimiliano Di Ventra}
\email{diventra@physics.ucsd.edu}
\affiliation{Department of
Physics, University of California, San Diego, La Jolla, California
92093-0319}

\begin{abstract}
The general Lagrange-Euler formalism for the three memory circuit
elements, namely, memristive, memcapacitive, and meminductive systems, is
introduced. In addition, {\it mutual meminductance}, i.e. mutual
inductance with a state depending on the past evolution of the
system, is defined. The Lagrange-Euler formalism for a general
circuit network, the related work-energy theorem, and the generalized Joule's first law are also
obtained. Examples of this formalism applied to specific circuits
are provided, and the corresponding Hamiltonian and its
quantization for the case of non-dissipative elements are
discussed. The notion of {\it memory quanta}, the quantum excitations of
the memory degrees of freedom, is presented. Specific examples are
used to show that the coupling between these quanta and the
well-known charge quanta can lead to a splitting of degenerate
levels and to other experimentally observable quantum effects.
\end{abstract}

\maketitle

\section{Introduction}
Circuit elements with memory, namely, memristive~\cite{chua71a,chua76a}, memcapacitive and meminductive~\cite{diventra09a} systems are attracting considerable attention in view of their application in diverse areas of science and technology, ranging from
solid-state memories~\cite{Green07a,Karg08a,Sawa08a} to neuromorphic circuits~\cite{pershin09c,jo10a,Choi09a,Lai10a,Alibart10a,fontana10a} and understanding of biological processes~\cite{pershin09b,Johnsen11a}.
The general axiomatic definition of memory elements considers any two fundamental circuit variables, $u(t)$ and $y(t)$ (i.e., current $I$, charge $q$, voltage $V$, or flux $\phi\equiv \int_{-\infty}^t V(t') dt'$) whose relation, the response $g$, depends also on
a set, $x=\{x_i\}$, of $n$ state variables describing
the internal state of the system. These variables could be, e.g., the spin polarization of the sample~\cite{pershin08a,wang09a} or the position of
oxygen vacancies in a thin film~\cite{strukov08a}. The resulting $n$-th order $u$-controlled memory circuit element is described by~\cite{diventra09a}
\begin{eqnarray}
y(t)&=&g\left(x,u,t \right)u(t) \label{Geq1}\\ \dot{x}&=&f\left(
x,u,t\right) \label{Geq2}
\end{eqnarray}
where $f$ is a continuous $n$-dimensional vector function. It is
assumed on physical grounds that, given an initial state $u(t=t_0)$
at time $t_0$, Eq.~(\ref{Geq2}) admits a unique
solution. If $u$ is the current and $y(t)$ is the voltage then Eqs. (\ref{Geq1}), (\ref{Geq2}) define memory resistive (memristive) systems. In this case $g$ is the {\em memristance} (for memory resistance). In memory capacitive (memcapacitive) systems, the
charge is related to the voltage so that $g$ is the {\em memcapacitance} (memory capacitance); while in memory inductive (meminductive) systems the flux is related to the current with $g$ the {\em meminductance} (memory inductance). These systems are characterized by a typical ``pinched hysteretic
loop'' in their constitutive variables when subject to a periodic input (with exceptions as discussed in Ref.~\onlinecite{pershin11a}). Indeed, we have recently argued that essentially all two-terminal electronic devices based on memory materials and systems, when subject to time-dependent perturbations, behave simply as - or as a combination of - memristors, memcapacitors and meminductors~\cite{ourrecentMT,diventra09a}. This unifying description is a source
of inspiration for novel digital and analog applications~\cite{pershin10c,pershin11a,pershin11d} and allows us to bridge apparently different areas of research.~\cite{diventra09a}

However, despite the wealth of applications and new ideas these
concepts have generated, it is nonetheless important to stress
that so far these memory elements have been discussed only within
their classical circuit theory definition, with quantum mechanics
entering at best in the microscopic parameters that determine the
state variables responsible for
memory\cite{strukov08a,diventra09a,Driscoll10b}. However, it seems
that these features are common at the nanoscale where the
dynamical properties of electrons and ions are likely to depend on
the history of the system, at least within certain time
scales~\cite{diventra09b,Maxbook}. Mindful of the trend towards
extreme miniaturization of devices of all sorts, it is thus
natural to ask whether true quantum effects can be associated with
the memory of these systems and which phenomena could emerge from
the quantization of memory elements. Of course, examples of memory
effects in quantum phenomena can be found in the specialized
literature (see, e.g., Ref. ~\onlinecite{Breuer2002a}). Here
instead, we want to provide a general framework of study of the
quantum excitations ({\it memory quanta}) associated to general
degrees of freedom that lead to memory in these systems.

We then first introduce the
general Lagrange-Euler formalism for these systems. This is the
non-trivial extension of the corresponding formalism for the
``standard'' circuit elements. Since it is well known that the
Lagrangian formulation of circuit elements offers great advantages
in the analysis of complex circuits~\cite{Devoret1997}, we expect
that this generalization would be of great value in itself.
Moreover, our work extends previous studies related to the formulation
of Lagrange and Routh equations for non-linear circuits involving
ideal memristors \cite{shragowitz88a} and to the port-Hamiltonian
modeling for the case of memristive components
\cite{jeltsema10a}. In the present context our work also sheds
light on the general relation between the internal degrees of
freedom that lead to memory and the constitutive variables - the
charge, current, voltage and flux - that define the different
elements. Along the way we also define {\it mutual meminductors}, namely
mutual inductors with memory, which add additional flexibility and hence
new functionalities to the field of memory elements.

We finally proceed to
quantize the corresponding equations in the standard way. This leads us to consider the memory excitations of these systems.
In this paper we consider only the quantization of non-dissipative elements, and we will devote a subsequent
paper to the discussion of quantum effects in dissipative memory elements. We will provide examples of
applications of the Lagrangian formalism to selected cases and discuss experimental conditions under which these memory quanta could be detected.

This paper is organized as follows. In Sec. \ref{sec:general} we introduce a general scheme of the approach. Sec. \ref{section2} is dedicated to the Lagrangian formulation of
memristive systems, while
Secs. \ref{section3} and~\ref{section4} deal with memcapacitive and meminductive systems, respectively. We then show how to write the Lagrangian (Sec. \ref{SEC:lagrangian-general-circuit}) and Hamiltonian (Sec. \ref{SEC:hf}) of a circuit of memory elements and also give the work-energy theorem and generalized Joule's first law for such a circuit. We introduce the
concept of memory quanta in Sec. \ref{section5} focusing on specific examples. Finally, in Sec. \ref{section6} we
report our conclusions.

\section{Lagrange approach} \label{sec:general}

In the Lagrange formalism, each memory circuit element is associated with $m+1$ degrees of freedom (one related to a circuit variable ($q$ or $\phi$) and $m$ to its internal state (generalized coordinates, $y_j$, $j=1,...,m$)). For  convenience, we define two multivariate vectors
\begin{eqnarray}
Y^q=(q,y_1,...,y_m) ,\label{eq:Yq} \\
Y^\phi=(\phi,y_1,...,y_m). \label{eq:Yphi}
\end{eqnarray}
We note that there are two (in some cases, however, one) internal state variables $x_i$ (entering Eqs. (\ref{Geq1}), (\ref{Geq2})) for each $y_j$. Quite generally then $x=\{ y,\dot{y} \}$ (with $y$ here not to be confused with the
output variable $y(t)$ in Eq.~(\ref{Geq1})).

A model of any particular memory circuit element consists of three components: the kinetic energy, $T$, the potential energy, $U$, and the dissipation potential ${\cal H}$. The $m+1$ Lagrange equations of motion are given by
\begin{equation}
\frac{\textnormal{d}}{\textnormal{d}t}\frac{\partial {\cal L}}{\partial \dot{Y}^\alpha_j}-\frac{\partial {\cal L}}{\partial Y^\alpha_j}=Q_{Y^\alpha_j},\label{eq:Lag}
\end{equation}
where ${\cal L}=T-U$ is the Lagrangian, $\alpha$ is $q$ or $\phi$, $j=0,...,m$, and the generalized dissipation force $Q_{Y^\alpha_j}$ is defined as
\begin{eqnarray}
Q_{Y^\alpha_j}=-\frac{\partial {\cal H}}{\partial \dot{Y}^\alpha_j} . \label{eq:dissip_force}
\end{eqnarray}
While, generally,  models of different memory circuit elements  involve similar terms related to internal degrees of freedom, the contribution from the circuit variable $q$ or $\phi$ is specific for each type of memory circuit element as presented in the Table \ref{table:overview}.

\begin{table}
\begin{tabular}{ | l | c | c | c | c |  }
  \hline
 $\;\;\;\;\;\;\;\;\;\;\;\;\;\;\;$ System type & variables & $T$ & $U$ & $\cal{H}$ \\
  \hline
  $V$-controlled memristive system & $Y^q$ &  & & ${\cal H}^M_V$ \\
   \hline
  $I$-controlled memristive system & $Y^\phi$ &  & & ${\cal H}^M_I$ \\
  \hline
    $V$-controlled memcapacitive system & $Y^q$ &  & $U^C_V$ &  \\
   \hline
  $q$-controlled memcapacitive system & $Y^\phi$ & $T^C_q$ & & \\
  \hline
    $\phi$-controlled meminductive system & $Y^q$ & $T^L_\phi$ & &  \\
   \hline
  $I$-controlled meminductive system & $Y^\phi$ &  & $U^L_I$ & \\
  \hline
\end{tabular}
\caption{General scheme of Lagrange description of memory circuit elements. Specific contributions listed in the columns $T$, $U$ and $\cal{H}$  are given by Eqs. (\ref{eq:HMV}), (\ref{CC3}), (\ref{eq:UCV}), (\ref{CCM1}), (\ref{eq:TLP}), (\ref{CCL2}).} \label{table:overview}
\end{table}

The kinetic energy $T$ may have a contribution describing the dynamics of internal degrees of freedom and a specific contribution according to Table \ref{table:overview}. The contribution from internal degrees of freedom, $\widetilde{T}$, can be written using symmetry arguments. First of all, since dissipative effects are not included in the kinetic energy, it is time-reversal invariant, and only even powers of $\dot{y}_i$ can exist. In order for the transformation
to canonical momenta be invertible, however, we must leave only
quadratic terms, and we find ${\widetilde T}=\sum_{ij} \widetilde{c}_{ij} \dot{y}_i\dot{y_j}$.
This form, being symmetric, can be diagonalized to give
\begin{equation}
T={\widetilde T}+T^{\beta}_{u}=\sum_i \frac{c_i \dot{\widetilde{y}}^2_i}{2}+T^{\beta}_{u},\label{eq:Tgen}
\end{equation}
where $c_i$  are real positive numbers to be determined
microscopically, and $T^{\beta}_{u}$ is the specific contribution, if it exists (see Table \ref{table:overview}) to the kinetic energy for $u$-controlled memory circuit element, $\beta=M,C$ or $L$.

The potential energy $U$ and dissipative potential $\cal H$ also include a specific contribution from Table \ref{table:overview} and contributions from internal degrees of freedom $\widetilde{U}$ and ${\widetilde{\cal H}}$:
\begin{eqnarray}
U=\widetilde{U} (y,u,t)+U^{\beta}_{u}, \label{eq:Ugen} \\
{\cal H}=\widetilde{\cal H} (y,\dot{y},u,t)+{\cal H}^{\beta}_{u}, \label{eq:Hgen}
\end{eqnarray}
where $y=\{y_i\}$. It is important to consider the control variable $u$ as an independent parameter that can be replaced by an (output) circuit variable (using, e.g., Eq. (\ref{Geq1})) only in the final equations of motion.

\section{Memristive systems}\label{section2}
There are two types of memristive systems: voltage-controlled and current-controlled ones~\cite{chua76a}.
From Eqs.~(\ref{Geq1}) and~(\ref{Geq2}), we define voltage-controlled memristive systems by the equations
\begin{eqnarray}
I_M(t)&=&R^{-1}\left(x,V_M,t \right)V_M(t), \label{eq1}\\
\dot{x}&=&f\left(x,V_M,t\right), \label{eq2}
\end{eqnarray}
where $V_M(t)$ and $I_M(t)=\dot{q}(t)$ denote the voltage and current across the
device, and $R$ is the memristance and its inverse is the memductance (for memory conductance). A current-controlled memristive system is such that
the resistance and the dynamics of state variables depend on the current \cite{chua76a,pershin11a}
\begin{eqnarray}
V_M(t)&=&R\left(x,I_M,t \right)I_M(t), \label{eq1-1}\\
\dot{x}&=&f\left(x,I_M,t\right). \label{eq1-2}
\end{eqnarray}

At this point we note that the above equations have been introduced to define a wide class of systems collectively called memristive~\cite{chua76a}, while the name memristor \cite{chua71a}
has been assigned to the ideal case of these equations, when $R$ depends only on the voltage (or current)
history. Although some authors use the
term memristor to represent any system that satisfies Eqs.~(\ref{eq1}),(\ref{eq2}) or (\ref{eq1-1}),(\ref{eq1-2}) we reserve this term for the ideal case only \cite{chua71a}. (We will also see in Sec.~\ref{mult} that such systems, like ideal memcapacitors and meminductors, require special care in the Lagrangian formulation.)
We also note that, often, current-controlled memristive systems can be redefined as voltage-controlled ones and vice-versa \cite{pershin11a}. In addition, according to Th\'{e}venin's theorem~\cite{Helmholtz1853a,thevenin1883a}, a voltage source $V(t)$ in series with
a resistance $R$ is equivalent to a current source $I(t)=V(t)/R$ with the same resistance in parallel. We could then choose to work with either one of these cases.
However, for completeness, in the following we will present the Lagrangian formalism for both voltage-controlled and current-controlled memristive systems.

\begin{figure}[t]
 \begin{center}
    \centerline{
    \mbox{(a)}
    \mbox{\includegraphics[width=2.00cm]{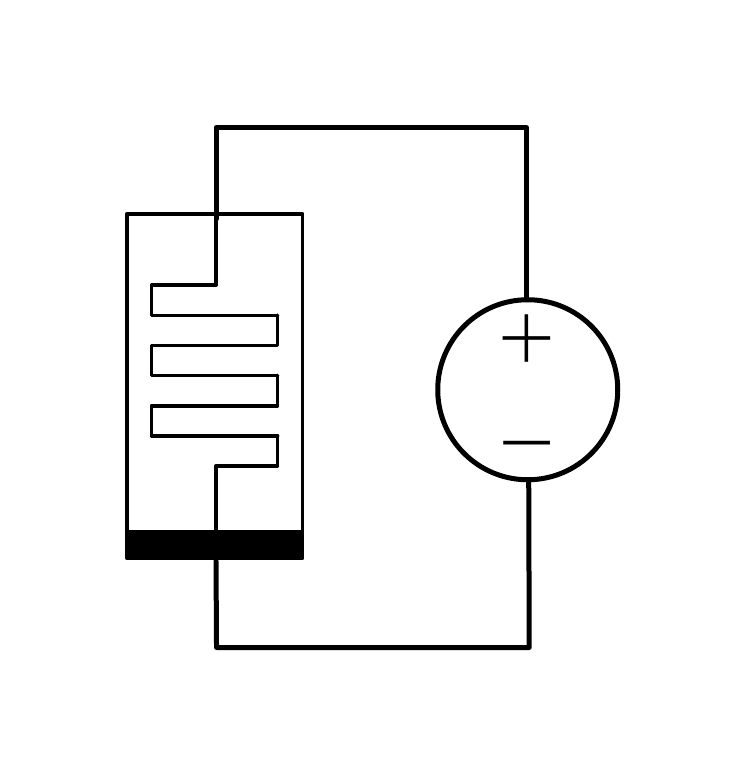}}
    \mbox{(b)}
    \mbox{\includegraphics[width=2.00cm]{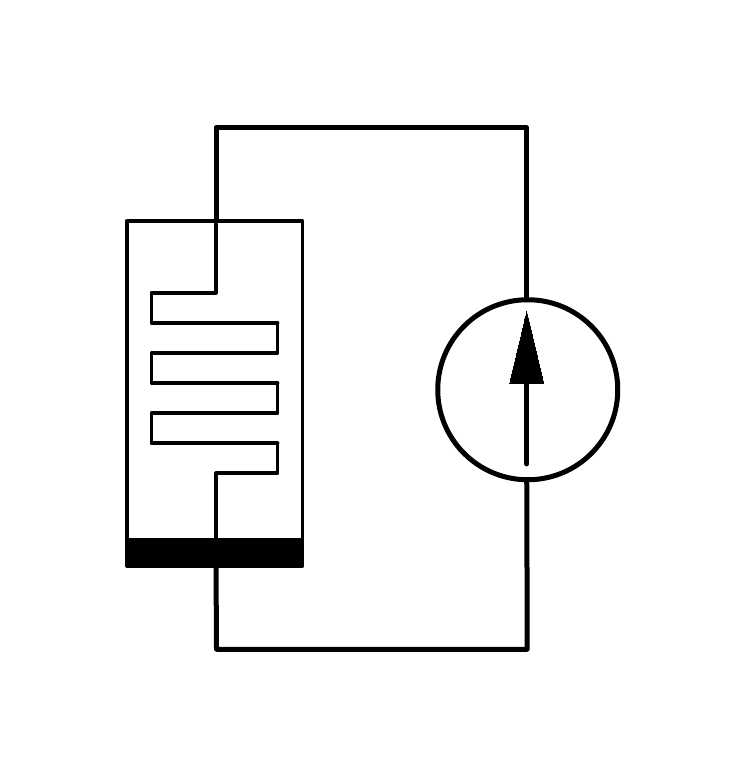}}
  }
\caption{\label{fig1} (a) Schematic of a
voltage-controlled memristive system connected to a time-dependent voltage source. (b) Schematic of a
current-controlled memristive system connected to a time-dependent current source.}
 \end{center}
\end{figure}

\subsection{Voltage-controlled systems}\label{seclagM}

We consider a voltage-controlled memristive system connected to a time-dependent voltage source $V(t)$ as shown in Fig. \ref{fig1}(a).
In addition to the term due to internal degrees of freedom discussed in Sec. \ref{sec:general},
the total potential energy contains the usual contribution from
the battery $-qV(t)$, with $q$ the charge that flows in the
circuit. There are many mechanisms for potential energy arising
from the state variables which are affected by the
applied bias --- an example of this is the change of state due to
electromigration (see, e.g., Ref.~\onlinecite{Maxbook}).
The total potential energy is thus given by
\begin{equation}
U=\widetilde U(y,V_M,t)-qV(t) \label{Upot}
\end{equation}
so that the Lagrangian is
\begin{equation}
{\cal L}=T-U=\sum_i \frac{c_i \dot{y}_i^2}{2}-\widetilde U(y,V_M,t)+qV(t).\label{Lageq}
\end{equation}
Here, $V_M$ is considered as an independent parameter.

As it is shown in Table \ref{table:overview}, the dissipation potential of voltage-controlled memristive systems includes a circuit variable contribution ${\cal H}^M_V$. We write it similarly to the well known  Rayleigh's ``dissipation potential'' (for a constant value resistor) of the type ${\cal H}=R\dot{q}^2/2$, which gives rise to a ``dissipation force''
$Q_{q}=-\nabla_{\dot{q}}{\cal H}=-R\dot{q}$~\cite{goldstein01a}. Specifically, we will use
\begin{equation}
{\cal H}^M_V= \frac{R\left(y,V_M,t \right) \dot{q}^2}{2}. \label{eq:HMV}
\end{equation}

At this point we stress that the memristance (as well as the memcapacitance and meminductance we will discuss later) may also depend on generalized velocities, $\dot{y}$, which are also included into $x$. This would simply
modify the Lagrange equations of motion without changing the overall formalism. To simplify the notation, however, we will not include this dependence explicitly here, and give an explicit example of this case in
Sec.~\ref{subseccl}.

For the total dissipation potential we write
\begin{equation}
{\cal H}=\frac{R\left(y,V_M,t \right) \dot{q}^2}{2}+{\widetilde{\cal H}}\left(y,\dot{y},V_M,t \right).
\label{totH}
\end{equation}
where the last term is to be determined phenomenologically or from a microscopic theory.

It is straightforward to show that the equation of motion (\ref{eq:Lag}) for $Y^q_0=q$  can be written as
\begin{equation}
V(t)\equiv V_M(t)=R(y,V_M,t)\,\dot{q}\label{firstLM}.
\end{equation}
This equation is of the type~(\ref{eq1}). The corresponding equations of motion for the state variables $x$ are
\begin{equation}
c_i\ddot{y}_i +\frac{\partial \widetilde{\cal H}\left(y,\dot{y},V_M,t \right)}{\partial \dot{y}_i}+\frac{\partial \widetilde U(y,V_M,t)}{\partial y_i}=0, \;\;\;
\label{secondLM}
\end{equation}
which show explicitly two possible physical origins of memristance
- due to a dissipative component and/or a potential energy
component.

Equation (\ref{secondLM}) can be rewritten as two first-order differential equations of the form (\ref{eq2}) considering both $y_i$ and $\dot{y}_i$ as internal state variables.
Moreover, in the final equations we can substitute $V_M$ by its expression in terms of the current $\dot{q}$. For this purpose, Eq. (\ref{eq1}) can be solved with respect to $V_M$. The same final procedure can also be used in the case of memcapacitive and meminductive systems
considered below.

\subsection{Current-controlled systems}\label{subsecccs}
As a simple example of a closed circuit with a current-controlled memristive system, we consider a source of current $I(t)$ connected to a memristive system (Fig. \ref{fig1}(b)). Here, as indicated in Eqs. (\ref{eq1-1}) and (\ref{eq1-2}), the output circuit variable is the voltage across the memristive system, $\dot{\phi}= V_M(t)$. That is why we use $Y^\phi$ set of variables in this case.
The kinetic energy, potential energy, and total dissipation potential in the Lagrangian formalism are now
\begin{eqnarray}
T&=&\half \sum\limits_i c_i \dot{y}_i^2,\label{CC1}\\
U&=&\widetilde{U}(y,I_M,t)-\phi I(t),\label{CC2}\\
{\cal H}&=&{\cal H}^M_I+\widetilde{\cal H}=
\frac{\dot{\phi}^2}{2R(y,I_M,t)}+\widetilde{\cal{H}}(y,\dot{y},I_M,t)\label{CC3},
\end{eqnarray}
where $-\phi I(t)$ is the battery term. Although not a necessary step (if their values are known), $R(y,I_M,t)$, $\widetilde{U}(y,I_M,t)$, $\widetilde{\cal{H}}(y,\dot{y},I_M,t)$ can be  obtained from $R(y,V_M,t)$, $\widetilde{U}(y,V_M,t)$ and $\widetilde{\cal{H}}(y,\dot{y},V_M,t)$, correspondingly. However, the solution may be multiple-valued in $I_M$ so that $R(y,\dot{y},I_M,t)$ may have multiple branches with the correct choice of branch depending on the history of the memristive system.

The equation of motion (EOM) for $\phi$ follows from Eq. (\ref{eq:Lag}) for $Y^\phi_0=\phi$, taking into account Eqs. (\ref{CC1}), (\ref{CC2}), and (\ref{CC3}), leading to
\begin{equation}
\dot{\phi}=R(y,I_M,t)I_M,
\end{equation}
which is just Eq. (\ref{eq1-1}). The EOM for $x$ is similarly found from Eq. (\ref{eq:Lag}), resulting in Eq. (\ref{secondLM}) except for the substitution of $V_M$ by $I_M$.
\subsection{Example}
Here we provide a specific physical example to clarify both the
formalism and the different terms that appear in
Eqs.~(\ref{firstLM}) and~(\ref{secondLM}). For this we consider a
thermistor, namely, a temperature-dependent resistor. The
memristive model of thermistor~\cite{chua76a,pershin11a} utilizes
a single internal state variable, the absolute temperature of
thermistor, $x=y=T_{therm}$, and can be formulated as first-order
voltage-controlled memristive system \cite{pershin11a}.
Mathematically, the Lagrangian model of thermistor involves the
following kinetic and potential energies and dissipation
potentials:
\begin{eqnarray}
T&=&0,\\
\widetilde U&=&0,\\
{\cal H}^M_V&=&\frac{R(y) \dot{q}^2}{2},\\
\widetilde {\cal H}&=&\dot{y}\left [
\frac{1}{2}C_h\dot{y}-\frac{V_M^2}{R(y)}-(T_{env}-y)\delta \right
],
\end{eqnarray}
where  $R(y)=R_0e^{\beta\left( 1/y-1/T_0\right)}$ is the temperature-dependent resistance, $R_0$ denotes the resistance at a certain temperature $T_0$, $\beta$ is a material-specific constant, $C_h$ is the heat capacitance, $\delta$ is the
dissipation constant of the thermistor \cite{chua76a}, and
$T_{env}$ is the background (environment) temperature.

Using Eq. (\ref{eq:Lag}) for a circuit consisting of a thermistor connected to a voltage source $V(t)$ (see Fig. \ref{fig1}(a)), we recover the equations of the memristive model of thermistor \cite{pershin11a}
\begin{eqnarray}
I&=&\left[R_0e^{\beta\left( 1/y-1/T_0\right)} \right]^{-1} V_M, \label{Itherm1}\\
C_h \frac{\textnormal{d}y}{\textnormal{d}t}&=&\left[R_0e^{\beta\left( 1/y-1/T_0\right)} \right]^{-1} V_M^2+(T_{env}-y)\delta . \;\;\;\label{Itherm2}
\end{eqnarray}
Note that although other forms of potential and kinetic energy terms could produce the same Eqs.~(\ref{Itherm1}) and~(\ref{Itherm2}) this particular one also satisfies the Joule's first law discussed in Sec.~\ref{SEC:hfWEThdiss}. This puts
severe constrains on the choice of Lagrangian.

\section{Memcapacitive systems} \label{section3}
\begin{figure}[t]
 \begin{center}
    \centerline{
    \mbox{(a)}
    \mbox{\includegraphics[width=2.00cm]{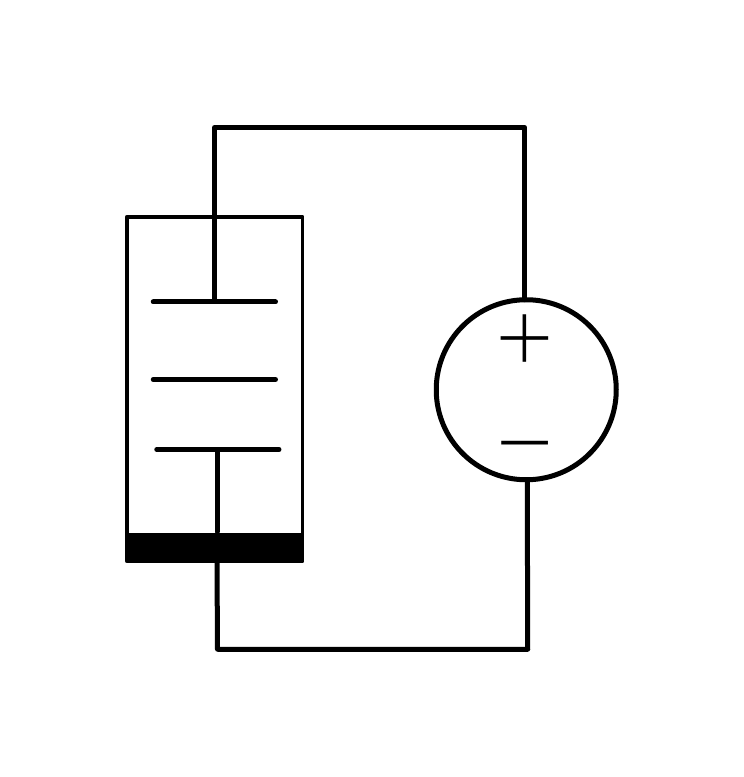}}
    \mbox{(b)}
    \mbox{\includegraphics[width=2.00cm]{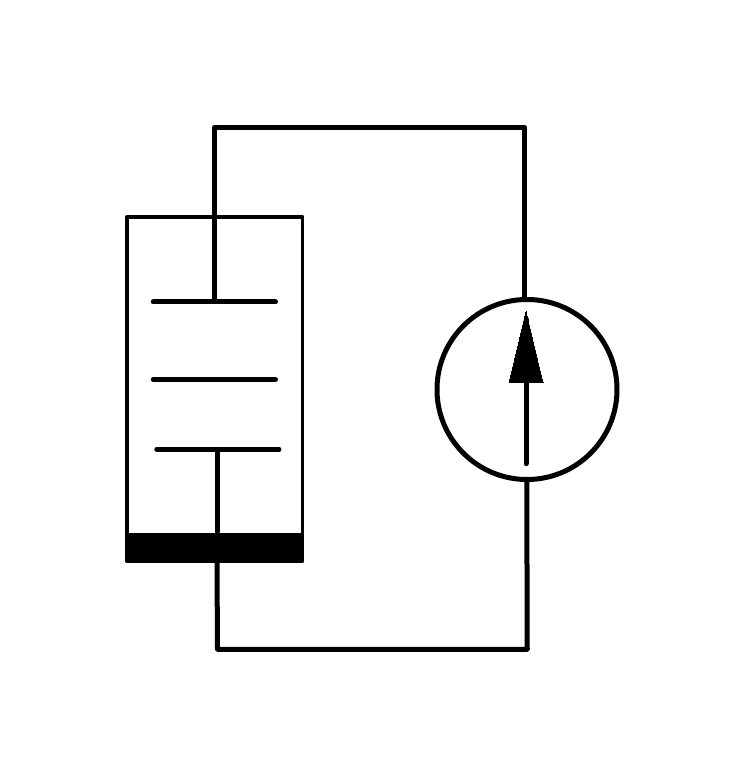}}
  }
\caption{\label{fig2} (a) Schematic of a
voltage-controlled memcapacitive system connected to a time-dependent voltage source. (b) Schematic of a
charge-controlled memcapacitive system connected to a time-dependent current source.}
 \end{center}
\end{figure}

We now consider memcapacitive systems~\cite{diventra09a} (Fig.~\ref{fig2}),
which---unlike memristive systems---store also energy.
In particular, voltage-controlled memcapacitive systems are defined by Eqs. (\ref{Geq1}) and~(\ref{Geq2}) with $u$ the voltage, $V_C(t)$, across the memcapacitive
system, and $y(t)$ the charge, $q_C(t)$, stored in the device, leading to
\begin{eqnarray}
q_C(t)&=&C\left(x,V_C,t \right)V_C(t) \label{Ceq1}, \\
\dot{x}&=&f\left( x,V_C,t\right) \label{Ceq2},
\end{eqnarray}
where $C$ is the memcapacitance.

As in memristive systems, the above equations define a large class of systems, with ideal memcapacitors
those for which the memcapacitance depends only on the voltage history (or for charge-controlled memcapacitive systems,
only on the charge history)~\cite{diventra09a}. In addition, it is often important to consider the energy
added to/removed from a memcapacitive system, namely the quantity $U_C=\int_{t_0}^{t}V_C(\tau)I(\tau)d\tau$ which helps understanding
whether a memcapacitive system is non-dissipative, dissipative, or active~\cite{diventra09a}. Of these, the non-dissipative and/or dissipative memcapacitive systems
are the most interesting for potential applications, and we will therefore focus here on these cases only.

A charge-controlled memcapacitive system is defined by the set of equations~\cite{diventra09a}
\begin{eqnarray}
V_C&=&C^{-1}(x,q_C,t)q_C(t)\label{ccms1},\\
\dot{x}&=&f(x,q_C,t)\label{ccms2}.
\end{eqnarray}

\subsection{Voltage-controlled systems}\label{seclagC}
The Lagrange model of voltage-controlled memcapacitive systems
is based on $Y^q$ set of variables (see Table \ref{table:overview}). The specific contribution from the
$q$ degree of freedom to the potential energy is
\begin{equation}
U^C_V=\frac{q^2}{2C(y,V_C,t)}. \label{eq:UCV}
\end{equation}
Taking into account a voltage source connected to the system (Fig. \ref{fig2}(a)),
the total potential energy is written as
\begin{equation}
U=\frac{q^2}{2C(y,V_C,t)}+\widetilde U(y,V_C,t)-qV(t)\label{Cenergy}.
\end{equation}
Consequently, the Lagrangian is given by
\begin{equation}
{\cal L}=T-U=\sum_i \frac{c_i \dot{y}_i^2}{2}-\frac{q^2}{2C(y,V_C,t)}-\widetilde U(y,V_C,t)+qV(t).\label{LageqC}
\end{equation}
The dissipative potential contains only the internal state variables contribution $\widetilde{\cal H}(y,\dot{y},V_C,t)$.

The Lagrange EOMs for voltage-controlled memcapacitive systems have the form
\begin{eqnarray}
\frac{q(t)}{C(y,V_C,t)}=V(t)\equiv V_C(t),\label{firstLC} \\
c_i\ddot{y}_i +\frac{\partial \widetilde{\cal H}\left(y,\dot{y},V_C,t
\right)}{\partial \dot{y}_i} \nonumber \\ + \frac{\partial
\widetilde U(y,V_C,t)}{\partial y_i}- \frac{V_C^2}{2}\frac{\partial
C(y,V_C,t)}{\partial y_i}=0, \label{secondLCC}
\end{eqnarray}
where in writing the last term in Eq.~(\ref{secondLCC}) we have made use of Eq.~(\ref{firstLC}).
Its clear that Eqs. (\ref{firstLC}), (\ref{secondLCC}) are of the form of Eqs. (\ref{Ceq1}), (\ref{Ceq2})
In fact, Eq. (\ref{secondLCC}) clearly shows that the
memory may arise from both a conservative potential contribution as well as a dissipative one.

Equation~(\ref{secondLCC}) describes an effective dynamical system and, together with Eq. (\ref{Ceq1}), tells us that, in the presence of a periodic input of frequency $\omega$,
charge dynamics can be out of phase with the voltage across the memcapacitive system.
Indeed, there might be delay in response of the internal state variables to the applied voltage leading to the above mentioned effect.
Experimentally, it can be seen as a pinched hysteresis loop in the $q-V_C$ plane \cite{diventra09a,pershin11a}.

\subsection{Charge-controlled systems}\label{subsecccc}

We consider a circuit consisting of a current source and a current-controlled memcapacitive system (Fig. \ref{fig2}(b)). Here, as seen in Eqs. (\ref{ccms1}) and (\ref{ccms2}), the circuit variable is the voltage across the memcapacitive system, $\dot{\phi}=V_C(t)$, instead of the current through it, $\dot{q}= I_C(t)$, as in voltage-controlled systems, c.f. Eqs. (\ref{Ceq1}), (\ref{Ceq2}).
 Consequently, our analysis should be based on the $Y^\phi$ set (Table \ref{table:overview}). The kinetic energy, potential energy, and total dissipation potential in the Lagrangian formalism are now
\begin{eqnarray}
T&=&\widetilde{T}+T^C_q=\half \sum\limits_i c_i \dot{y}_i^2+\half C(y,q_C,t) \dot{\phi}^2\label{CCM1},\\
U&=&\widetilde U(y,q_C,t)-\phi I(t)\label{CCM2},\\
\mathcal{H}&=&\widetilde{\cal H}(y,\dot{y},q_C,t)\label{CCM3}.
\end{eqnarray}
The EOM for $\phi$ is derived by applying Eq. (\ref{eq:Lag}) to Eqs. (\ref{CCM1}), (\ref{CCM2}), and (\ref{CCM3}), leading to Eq. (\ref{ccms1}).
The EOM for $x$ is similarly obtained and results in Eq. (\ref{secondLCC}) except for the substitution of $V_C$ by  $q_C$.

\subsection{Example}
As example of a voltage-controlled memcapacitive system we consider a parallel-plate capacitor
with elastically suspended upper plate and a fixed lower plate \cite{pershin11a}. When charge is added to the plate, the separation between plates changes as oppositely charged plates experience an attractive interaction. The internal degree of freedom of the elastic memcapacitive system is the position of the upper plate $y$ measured from an equilibrium uncharged plate separation, $d_0$.  The Lagrange model of elastic memcapacitive system connected to a voltage source consists of the following kinetic and potential energies and dissipation potentials:
\begin{eqnarray}
T&=&\frac{m\dot{y}^2}{2}, \label{el1} \\
U&=&U^C_q+\widetilde U=\frac{q^2}{2C(y)}+\frac{ky^2}{2}-qV(t), \\
{\cal H}&=&\tilde{\cal H}= \frac{\gamma m\dot{y}^2}{2},\label{el4}
\end{eqnarray}
supplemented by the expression for the
memcapacitance, $C(y)=C_0/\left(1+y/d_0\right) $. Here, $m$ is the
mass of the upper plate, $\gamma$ is a damping coefficient
representing dissipation of the elastic oscillations, $k$ is the
spring constant, $C_0=\varepsilon S/d_0$ is the equilibrium value
of capacitance, $S$ is the plate area, and $\varepsilon$ the permittivity of the medium.

The first equation of motion is of the form of Eq. (\ref{firstLC}). The second equation (\ref{Ceq2}) is obtained substituting Eqs. (\ref{el1})-(\ref{el4}) into Eq. (\ref{secondLCC}). Explicitly, we obtain the classical harmonic oscillator equation including damping and driving terms:
\begin{equation}
\frac{\textnormal{d}^2y}{\textnormal{d}t^2}+\gamma
\frac{\textnormal{d} y}{\textnormal{d}t}+\omega_0^2
y+\frac{V_C^2}{2md_0}\frac{C_0}{\left(1+y/d_0\right)^2}=0. \label{el7}
\end{equation}
Here, $\omega_0=\sqrt{k/m}$. To emphasize the similarity of Eq. (\ref{el7}) with Eq. (\ref{Ceq2}) we note that Eq. (\ref{el7}) can be written as two first-order differential equations and the internal state variables are $x_1=y$ and $x_2=\dot{y}$.

\section{Meminductive systems} \label{section4}
\begin{figure}[t]
 \begin{center}
    \centerline{
    \mbox{(a)}
    \mbox{\includegraphics[width=2.00cm]{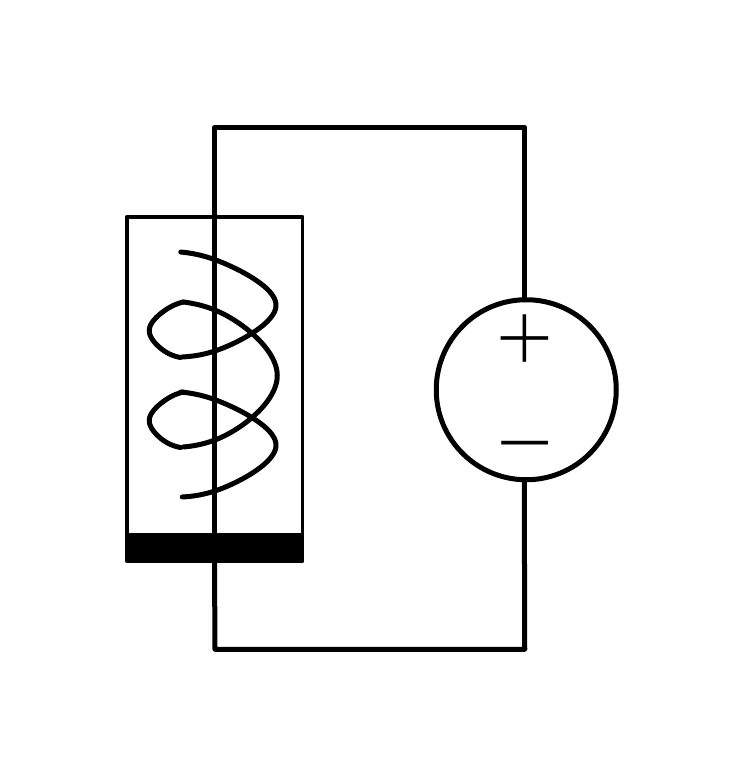}}
    \mbox{(b)}
    \mbox{\includegraphics[width=2.00cm]{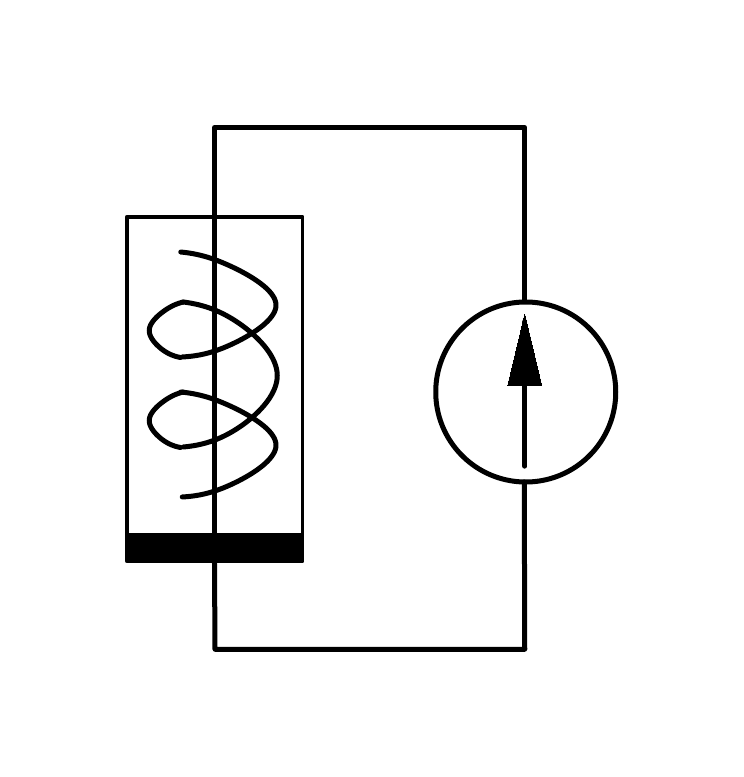}}
  }
\caption{\label{fig3}  (a) Schematic of
flux-controlled meminductive system connected to a time-dependent voltage source. (b) Schematic of a
current-controlled meminductive system connected to a time-dependent current source.}
 \end{center}
\end{figure}
Let us finally consider meminductive systems~\cite{diventra09a} (Fig.~\ref{fig3}). A flux-controlled meminductive system satisfies the relations~\cite{diventra09a}
\begin{eqnarray}
I_L&=&L^{-1}\left(x,\phi_L,t \right)\phi_L(t) \label{LLeq1} \\
\dot{x}&=&f\left( x,\phi_L,t\right) \label{LLeq2}
\end{eqnarray}
with $L^{-1}$ the inverse meminductance. A current-controlled meminductive system is defined by the set of equations~\cite{diventra09a}
\begin{eqnarray}
\phi_L&=&L(x,I_L,t)I_L(t)\label{ccls1},\\
\dot{x}&=&f(x,I_L,t)\label{ccls2}.
\end{eqnarray}
As in the case of memcapacitive systems, meminductive elements
may represent non-dissipative, dissipative, or active devices. We are interested only in the first two types since
they are the most important for technological applications.

\subsection{Flux-controlled systems}\label{seclagL}
Consider a circuit composed of a voltage source connected to a meminductive system as in Fig. \ref{fig3}(a). The circuit degree of freedom $q$ in
flux-controlled meminductive systems is taken into account by the following contribution to the kinetic energy
\begin{equation}
T^L_\phi=\frac{L(y,\phi_L,t)\dot{q}^2}{2}. \label{eq:TLP}
\end{equation}
The contribution to $T$, $U$ and $\cal H$ from internal state degrees of freedom are written in the general form (Eqs. (\ref{eq:Tgen}),(\ref{eq:Ugen}),(\ref{eq:Hgen})). Consequently, taking also a voltage source in Fig. \ref{fig3}(a) into account,
the Lagrangian and dissipative potential are written as
\begin{eqnarray}
{\cal L}&=&\sum_i \frac{c_i \dot{y}_i^2}{2}+L(y,\phi_L,t)\frac{\dot{q}^2}{2}-\widetilde U(y,\phi_L,t)+qV(t),\;\;\;\;\;\;\;\label{LageL} \\
{\cal H}&=&\widetilde{\cal H} \left( y,\dot{y},\phi_L,t \right).
\end{eqnarray}

The EOM for the $q$ degree of freedom is
\begin{equation}
\frac{d(L(y,\phi_L,t) \dot{q})}{dt} =V(t)\equiv V_L(t).
\end{equation}
Integrating this equation in time assuming that $\phi(t=-\infty)=0$ we find
\begin{equation}
L(y,\phi_L,t)
\dot{q}=\int_{-\infty}^{t}dt'\,V_L(t')=\phi_L(t),\label{FL3}
\end{equation}
which is Eq.~(\ref{LLeq1}).

The EOMs for the state variables are written as
\begin{eqnarray}
c_i\ddot{y}_i &+& \frac{\partial \widetilde{H} \left( y,\dot{y},\phi_L,t
\right)}{\partial \dot{y}_i} \nonumber \\ &+& \frac{\partial \widetilde
U(y,\phi_L,t)}{\partial y_i}- \frac{\phi_L^2}{2L^2}\frac{\partial
L(y,\phi_L,t)}{\partial y_i}=0, \label{secondLC}
\end{eqnarray}
which again, since $x=\{y,\dot{y}\}$, can be written in the form of Eq.~(\ref{LLeq2}).

\subsection{Current-controlled systems}\label{subseccl}
 When one considers circuits involving current-controlled meminductive systems (and current sources instead of voltage sources), one should employ the $Y^\phi$ set of variables (see Table \ref{table:overview}). Let us then consider a simple circuit composed of a current source directly connected to a meminductive system (Fig. \ref{fig3}(b)).
The contribution from the circuit degree of freedom $\phi$ comes from the potential energy $U^L_I$ term.
The kinetic energy, potential energy, and total dissipation potential in the Lagrangian formalism are now
\begin{eqnarray}
T&=&\widetilde{T}=\half \sum\limits_i c_i \dot{y}_i^2\label{CCL1},\\
U&=&\widetilde{U}+U^L_I-\phi I=\widetilde{U}(y,I_L,t)+\frac{\phi^2}{2L(y,I_L,t)}-\phi I(t), \;\;\;\;\;\;\;\;\label{CCL2}\\
\mathcal{H}&=&\widetilde{\cal{H}}(y,\dot{y},I_L,t)\label{CCL3}.
\end{eqnarray}
The EOM for $\phi$ is the same as Eq. (\ref{ccls1}).
The EOMs for $y_i$ are similarly derived and result in Eq. (\ref{secondLC}) except for the substitution of $\phi_L$ by $I_L$.

\subsection{Example}
We here provide an instructive example of an effective meminductive system consisting of an
LCR contour inductively coupled to an inductor (Fig.~\ref{fig4}). In this scheme, the two inductors $L_1$ and $L_2$
interact with each other magnetically. From the point of view of the voltage source $V(t)$, the total system can be
seen as a second-order flux-controlled meminductive system described by the general equations (\ref{LLeq1})--(\ref{LLeq2}).
The charge on the capacitor C and the current through the inductor
$L_2$ play the role of internal state variables. It is convenient
to select $y=q_C$. Consequently, $I_{L_2}=\dot{y}$.

We start by considering the circuit presented in Fig. \ref{fig4} using the Lagrange formalism for usual circuit elements.
The circuit is described by:
\begin{figure}[t]
 \begin{center}
    \includegraphics[width=5.69cm]{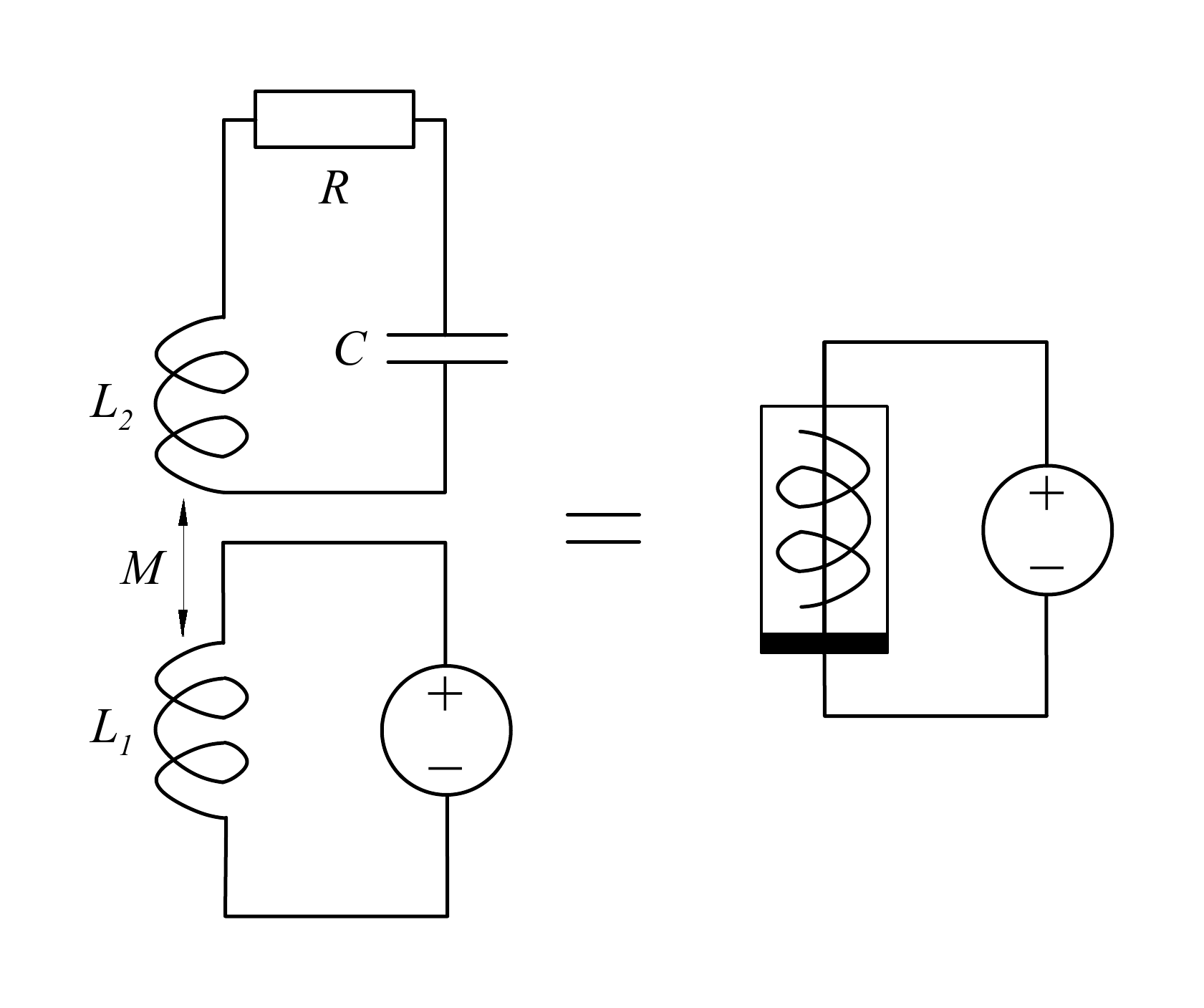}
\caption{\label{fig4} Flux-controlled meminductive system based on the inductive coupling of a coil $L_1$ with a LCR contour. Here, the
mutual inductance $M$ is equal to $k\sqrt{L_1L_2}$, where $0 \leq k \leq 1$ is the coupling coefficient.}
\end{center}
\end{figure}
\begin{eqnarray}
T&=&\frac{1}{2}L_1\dot{q}^2+\frac{1}{2}L_2\dot{y}^2+M\dot{y}\dot{q}, \label{mnk1} \\
U&=&\frac{y^2}{2C}-qV(t), \\
{\cal H}_q&=&0, \\
{\cal H}_x&=& \frac{R\dot{y}^2}{2}.\label{mnk4}
\end{eqnarray}
The EOMs for $x$ and $q$ are then found to be
\begin{equation}
L_1\ddot{q}+M\ddot{y}-V(t)=0, \label{mi1}
\end{equation}
\begin{equation}
\left( L_2-\frac{M^2}{L_1}\right
)\ddot{y}+\frac{M}{L_1}V(t)+\frac{y}{C}+R\dot{y}=0.\label{mi2}
\end{equation}
One can easily verify that Eqs. (\ref{mi1}) and (\ref{mi2})
describe the electric circuit from Fig. \ref{fig4}.

Next, integrating Eq. (\ref{mi1}) (the constant of integration is taken to be zero), we can rewrite it in the form
\begin{equation}
\phi(t)=\frac{L_1 \phi(t)}{\phi(t)-M\dot{y}}\dot{q}(t)\equiv L(\dot{y},\phi(t)) \dot{q}(t),
\end{equation}
which shows that the meminductance $L$ depends on the generalized velocity $x=\dot{y}$.

\subsection{Mutual meminductance}
After the generalization of self-inductance to meminductance, one wonders if mutual-inductance can be generalized to memory situations as well. We consider two coupled inductors as in Fig. \ref{fig4},
but now assume the mutual inductance to have memory. Here, we want to describe that part of the memory that cannot be included in two (self-)meminductive systems. This memory can be stored in the medium between the inductors with a state affected by the two magnetic fluxes of the inductors.
It could also be stored in the geometry of the system by having, e.g., two elastic coils that can either attract or repel each other. Since the memory mechanism does not belong solely to one inductor, the relation $M=k\sqrt{L_1L_2}$, applicable to mutual inductance of two coils, does not apply for mutual meminductance: $k$ is not generally a constant independent of $L_1$, $L_2$, $x$ and possibly some other parameters.

\begin{figure}[t]
 \begin{center}
 \includegraphics[width=2.78cm]{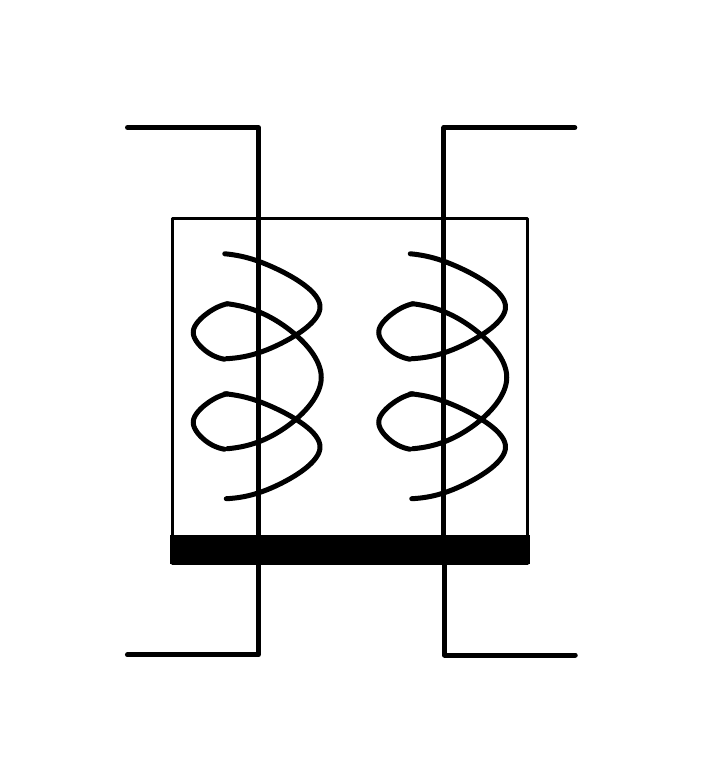}
\caption{\label{fig10}  Symbol for a mutual meminductor.}
 \end{center}
\end{figure}

In analogy with Eqs. (\ref{LLeq1}) and (\ref{LLeq2}), we then define a flux-controlled \emph{mutual meminductive system} via the following set of equations:
\begin{eqnarray}
\dot{q}_1&=&M^{-1}(x,\phi_{M1},\phi_{M2},t)\phi_{M2},\\
\dot{q}_2&=&M^{-1}(x,\phi_{M1},\phi_{M2},t)\phi_{M1},\\
\dot{x}&=&f(x,\phi_{M1},\phi_{M2},t),
\end{eqnarray}
where $M(x,\phi_{M1},\phi_{M2},t)$ is the mutual meminductance, $\phi_{M1}$ is the magnetic flux defined by $\phi_{M1}=\int_{-\infty}^t V_{M1}(t')dt'$, $V_{M1}(t)$ is the voltage on the first inductor
 ($\phi_{M2}$ and $V_{2M}(t)$ are similarly defined), and $\dot{q}_1$ and $\dot{q}_2$ are the currents in the first and second inductors
 respectively. The circuit symbol we propose for this element is
 shown in Fig. \ref{fig10}.

Regarding the Lagrangian formulation, the additions to the kinetic
energy, potential energy and dissipation potential as a result of
introducing this memory element are
\begin{eqnarray}
T&=&\half \sum\limits_i c_i \dot{x}_i^2+M(x,\phi_{M1},\phi_{M2},t)\dot{q}_1\dot{q}_2,\\
U&=&\widetilde{U}(x,\phi_{M1},\phi_{M2},t),\\
\mathcal{H}&=&\mathcal{H}_x(x,\dot{x},\phi_{M1},\phi_{M2},t).
\end{eqnarray}

The corresponding current-controlled mutual meminductive systems are instead defined by the set of equations
\begin{eqnarray}
\phi_{M1} &=&M(x,I_{M1},I_{M2},t) I_{M2},\\
\phi_{M2} &=&M(x,I_{M1},I_{M2},t) I_{M1},\\
\dot{x}&=&f(x,I_{M1},I_{M2},t),
\end{eqnarray}
where $I_{Mi}$ is the current in the $i$-th inductor, and $M(x,I_{M1},I_{M2},t)$ is the mutual inductance that can be obtained by plugging in $\phi_{M1}(I_{M2})$ and $\phi_{M2}(I_{M1})$ in $M(x,\phi_{M1},\phi_{M2},t)$. The Lagrangian formulation of this system is given by
\begin{eqnarray}
T&=&\half \sum\limits_i c_i \dot{x}_i^2,\\
U&=&\widetilde{U}(x,M I_{M2}, M I_{M1} ,t)+\frac{\phi_{M1}\phi_{M2}}{M(x,I_{M1},I_{M2},t)},\\
\mathcal{H}&=&\mathcal{H}_x(x,\dot{x},M I_{M2}, M I_{M1},t),
\end{eqnarray}
where both $\widetilde{U}$ and $\mathcal{H}_x$ are the same functions as defined in the above flux-controlled case.

\section{Lagrangian of a general circuit}\label{SEC:lagrangian-general-circuit}
We now have all the ingredients to write down the Lagrangian for a
general circuit network composed of an arbitrary combination of
memristive, memcapacitive and meminductive systems and their standard
counterparts. These circuits may be powered by an arbitrary set of
voltage sources $V_k(t)$ (Sec. \ref{subseccvce}), for which the fluxes
$\phi_k(t)$ are defined as, e.g., in Eq.~(\ref{FL3}), or by a set of current sources $I_k(t)$ (Sec. \ref{SUBSEC:ccce}). When both voltage sources and current sources are present, one can convert the latter to the former using the Th\'{e}venin's theorem, or the former to the latter using the Norton's theorem~\cite{NortTheorem}, thus ensuring only one type of power source is present. Below we briefly outline the recipe to write the Lagrangian of a general circuit for both cases.

\subsection{Circuits with voltage sources}\label{subseccvce}
A general electronic circuit powered by voltage sources can be described
as a combination of $l$ indivisible loops, i.e., ones
that do not contain internal loops. Within
the $j$-th ($j=1,\dots,l$) loop one should consider
the charge $q_j$ as the circuit variable and take into account
generalized coordinates of elements involved in this loop.
 For simplicity, we rename the generalized coordinates for the
whole circuit as $y_i$ ($i=1,\dots,k$).

The current in each branch of the circuit is
the sum of contributions from indivisible loops it belongs
to. Using this fact, we can write the Lagrangian for each
element in the branch. The element's Lagrangian is taken in the
voltage-controlled form for memristive and memcapacitive systems and in
the flux-controlled form for meminductive ones.

The sum of the Lagrangians of individual elements of the circuit
gives the circuit's Lagrangian,
while the sum of the dissipation potentials gives the circuit's dissipation potential.
The circuit's Lagrangian and dissipation potential depend
on $q_j$, $\dot{q}_j$, $y_i$ and $\dot{y}_i$ and result in
$l+k$ EOMs. The EOM obtained for $q_j$ gives Kirchhoff's
voltage law (KVL) for the $j$-th loop because of the linearity of the
Euler-Lagrange equations, and because each component in the loop
was shown above to give the correct voltage term. Kirchhoff's
current law (KCL), on the other hand, is automatically satisfied by the
choice of loop current variables.

\subsection{Circuits with current sources}\label{SUBSEC:ccce}
When a circuit is powered by current sources, the circuit variable is
the flux $\phi_j$ in the $j$-th ($j=1,\dots,l$) junction, while $\dot{\phi}_j$ is the electric potential at the junction.
 Using this definition, the flux or voltage across each element
in the network can be found via the difference of the fluxes or potentials in the junctions at its ends, enabling one to write the element's Lagrangian and dissipation potential in the current-controlled formalism for memristive and meminductive systems and in the charge-controlled formalism for memcapacitive ones. As in the voltage-controlled case, the circuit's Lagrangian or dissipation potential is the sum of the circuit element's Lagrangians or dissipation potentials, respectively.

If we denote again the internal degrees of freedom of the whole circuit as $y_i$ ($i=1,\dots,k$), we have a circuit's Lagrangian and dissipation potential that depend on $\phi_j$, $\dot{\phi}_j$, $y_i$ and $\dot{y}_i$ and result in
$l+k$ EOMs. The EOM obtained for $\phi_j$ gives KCL for the $j$-th junction due to the linearity of the
Euler-Lagrange equations, and because each element ending on the junction was shown above to give the correct current term. KVL, on the other hand, is automatically satisfied by the choice of junction potential variables.

This formalism has a complementary nature and can be viewed as the \emph{dual formalism} to that for circuits with voltage sources.
This conclusion will be reinforced in Sec. \ref{SEC:hf}, where we will show that the canonically conjugate momenta of the voltage-controlled and current-controlled formalisms to be fluxes and charges, respectively.

\subsection{Lagrangian multipliers}\label{mult}
When a circuit, voltage-controlled or current-controlled, has
additional constraints---missing from the EOMs---relating state
variables to circuit variables, the form of their dependence
should be added to the Lagrangian. This is achieved by the method
of Lagrange multipliers. In particular, the Lagrange multipliers
are convenient for describing ideal memory circuit elements such
as the ideal memristor~\cite{pershin11a}, in which the state
variable $y$ equals the charge $q$ flowing through the device.
Examples for such constraints appear in Subsec. \ref{examplesLM}.

\subsection{Examples}\label{examplesLM}
\begin{figure}[t]
 \begin{center}
    \includegraphics[width=8.00cm]{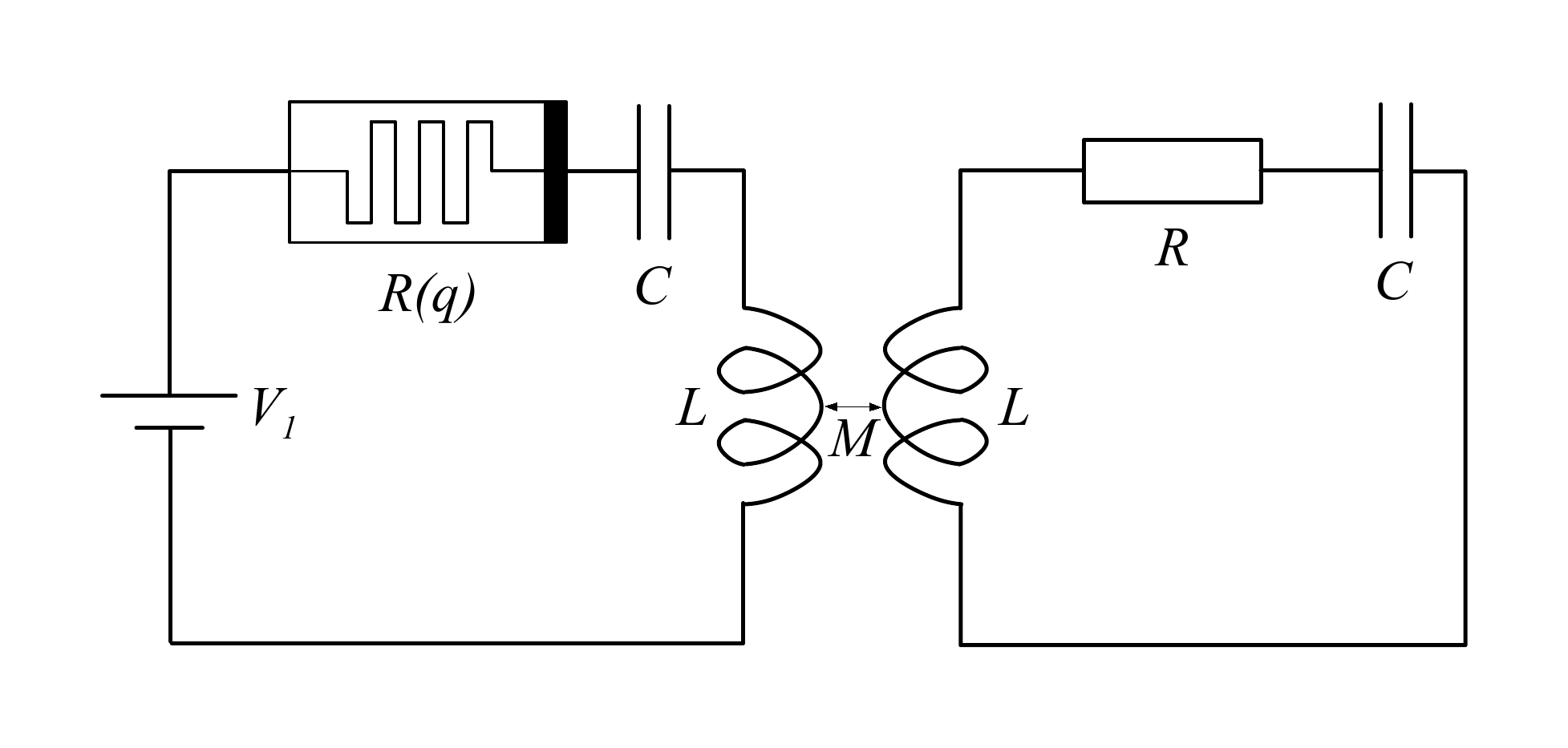}
\caption{\label{fig5} Schematic of flux-controlled
inductively-coupled charging circuits with a DC voltage source}
\end{center}
\end{figure}
\begin{figure}[t]
 \begin{center}
    \mbox{\includegraphics[width=8cm]{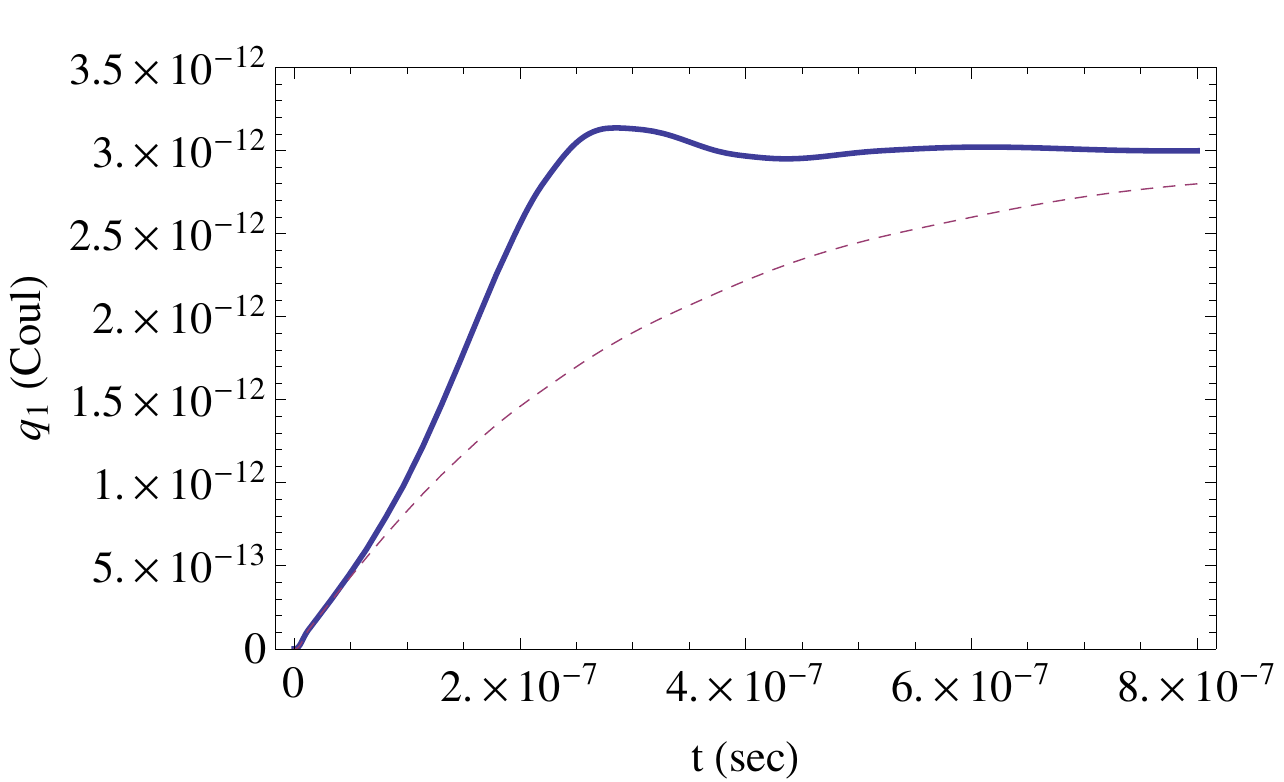}}
    \mbox{\includegraphics[width=8cm]{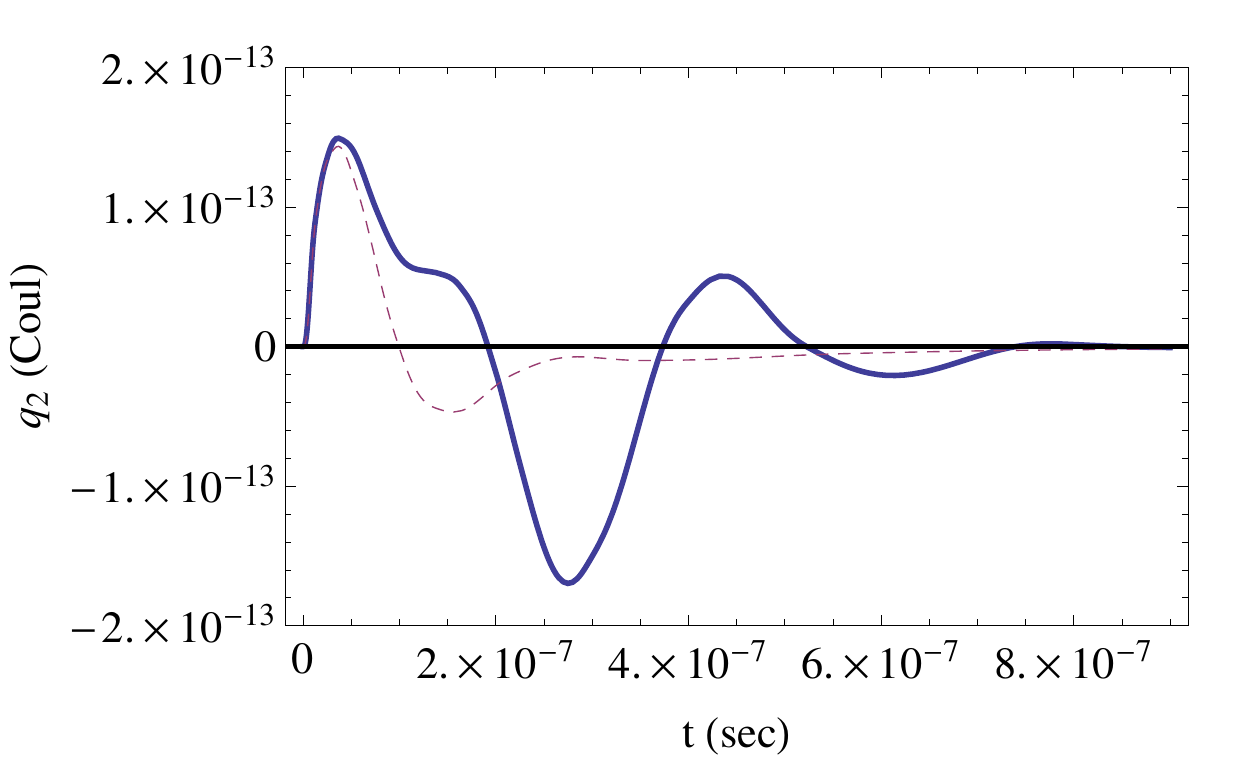}}
\caption{\label{fig6} (Color online) Graphs of $q_1(t)$ (top) and
$q_2(t)$ (bottom) for the system described in Fig. $\ref{fig5}$.
The solid line corresponds to the behavior with the memristor and
the dashed line to the behavior with the memristor replaced by a
normal resistor of resistance $R_{\mathrm{off}}$. The parameters
used were $V_1=1$V, $R=10$k$\Omega$,
$R_{\mathrm{off}}=100$k$\Omega$, $R_{\mathrm{on}}=100\Omega$,
$C=3$pF, $M=L=0.3$mH, $q_0=10^{-12}$C. The initial conditions
were set to no charge or current in any of the circuit elements.}
 \end{center}
\end{figure}

Consider two inductively coupled $RC$ circuits as shown in Fig.
\ref{fig5}. Initially, both capacitors are not charged and there
are no currents in the circuits. The circuits are coupled via
mutual inductance $M$, which results in periodic charging and
discharging of the right side capacitor as will be seen below. We
take a memristor with $R(y)$ ($y=q$) qualitatively similar to the
one fitted recently to experiments on TiO$_2$ thin
films\cite{strukov08a}, namely of the form
\begin{equation}
R(y)=R_{\mathrm{on}}+\frac{R_{\mathrm{off}}-R_{\mathrm{on}}}{1+y^2/q_0^2},
\end{equation}
where $R_{\mathrm{on}}$, $R_{\mathrm{off}}$ and $q_0$ are
parameters defined for each memristor. The resistance is seen to
decrease from $R_{off}$ to $R_{on}$ as charge flows through the
memristor. Denoting the left and right loop charges as $q_1$ and
$q_2$, respectively, and applying the results of Sections
\ref{section2}, \ref{section3}, and \ref{section4}, we obtain the
following Lagrangian and dissipation potentials for the network,
\begin{eqnarray}
\mathcal{L}&=&\frac{L}{2}
(\dot{q}_1^2+\dot{q}_2^2)-M\dot{q}_1\dot{q}_2+q_1V_1-\frac{q_1^2+q_2^2}{2C}\nonumber\\
&&-\lambda(y-q_1),\label{holo}\\
\mathcal{H}_q&=&\half R(y)\dot{q}_1^2+\half R\dot{q}_2^2,\\
\mathcal{H}_x&=&0,
\end{eqnarray}
where $\lambda$ is the Lagrange multiplier corresponding to the
circuit holonomic constraint, $y=q_1$ (a different constrain would
lead to a different corresponding term in Eq.~(\ref{holo})). These
expressions lead to the following EOMs for the total system
\begin{eqnarray}
L\ddot{q}_1-M\ddot{q}_2+R(q_1)\dot{q}_1+\frac{q_1}{C}-V_1&=&0,\\
L\ddot{q}_2-M\ddot{q}_1+R\dot{q}_2+\frac{q_2}{C}&=&0,
\end{eqnarray}
where the EOM for $y$, giving $\lambda=0$, and the EOM for $\lambda$, giving
$y=q_1$, were substituted. The solution of these EOMs for certain
values of the parameters is shown in Fig. $\ref{fig6}$. We note
that the insertion of the memristor produces an almost constant
current instead of an exponentially decreasing one in the left
loop of the circuit. The stabilization of the current is achieved
by the decline in the characteristic charging time $R(q_1)C$ as
the capacitor is charged. The memristor also modifies the exchange
of energy between the two circuits, giving pronounced oscillations
in the charge of the right loop of the circuit, which are absent
when the memristor is substituted with a normal resistor.

\begin{figure}[t]
 \begin{center}
    \includegraphics[width=8.00cm]{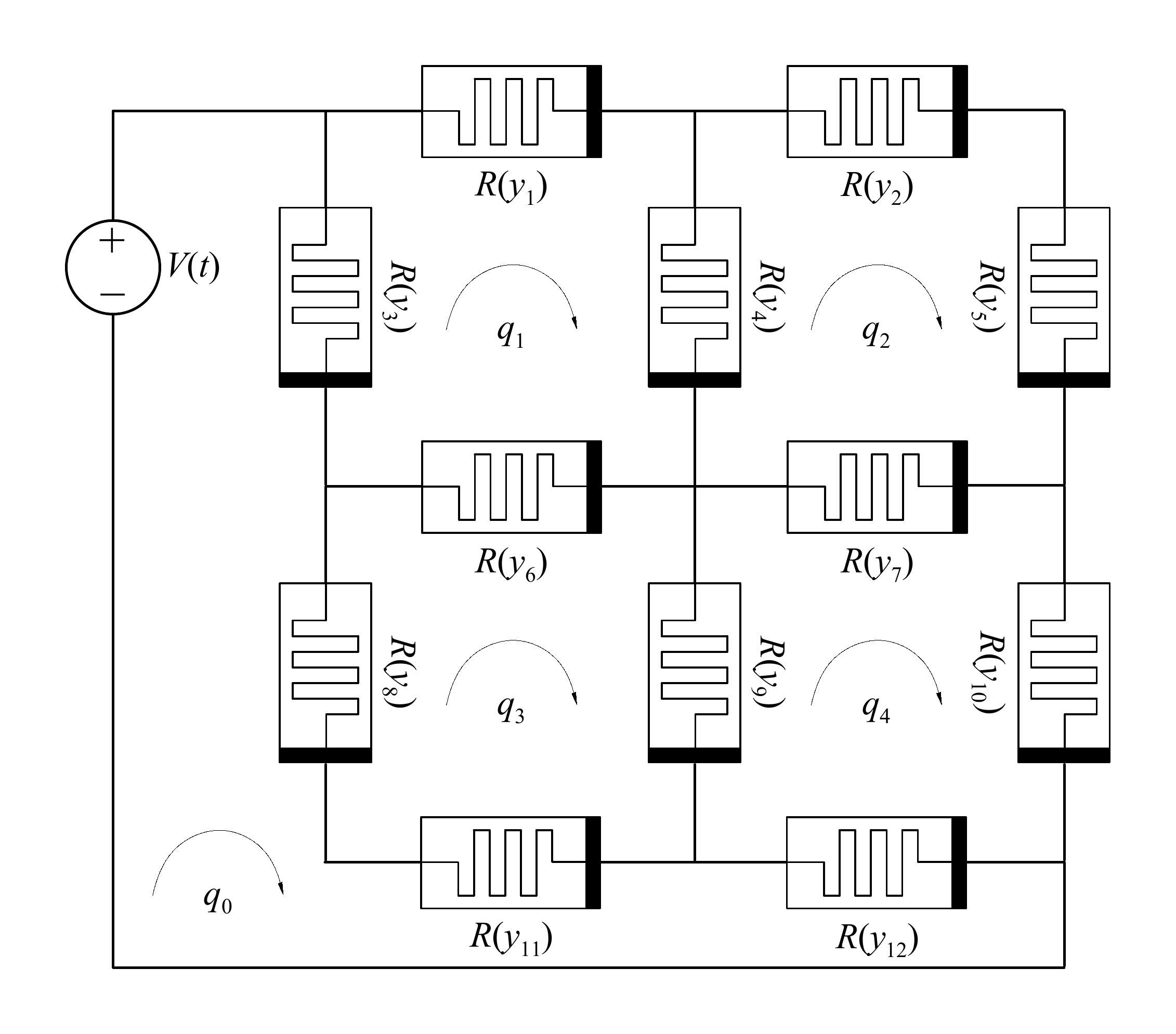}
\caption{\label{fig9} Schematic of a voltage-controlled $3\times
3$ network of memristive systems with a voltage source.}
\end{center}
\end{figure}
As a second example, consider the circuit shown in Fig.
\ref{fig9}. The circuit consists of a $3\times 3$ network of
memristors connected to a voltage source. Each memristor has
resistance $R(y_i)$, where $y_i$ is the cumulative charge that
flows through the memristor. The indivisible loops charges are
denoted by $q_i$. The EOMs can be readily obtained from the
Lagrangian and dissipation potential of the circuit that read
\begin{widetext}
\vspace{-20pt}
\begin{eqnarray}
\mathcal{L}&=&qV(t)-\lambda_1(y_1-q_1)-\lambda_2(y_2-q_2)-\lambda_3(y_3-q_0+q_1)-\lambda_4(y_4-q_1+q_2)-\lambda_5(y_5-q_2)\nonumber\\
&&-\lambda_6(y_6-q_3+q_1)-\lambda_7(y_7-q_4+q_2)-\lambda_8(y_8-q_0+q_3)-\lambda_9(y_9-q_3+q_4)-\lambda_{10}(y_{10}-q_4)\nonumber\\
&&-\lambda_{11}(y_{11}-q_0+q_3)-\lambda_{12}(y_{12}-q_0+q_4),\\
\mathcal{H}&=&\frac{1}{2}\{R(y_1)\dot{q}_1^2+R(y_2)\dot{q}_2^2+R(y_3)(\dot{q}_0-\dot{q}_1)^2+R(y_4)(\dot{q}_1-\dot{q}_2)^2+R(y_5)\dot{q}_2^2+R(y_6)(\dot{q}_3-\dot{q}_1)^2+R(y_7)(\dot{q}_4-\dot{q}_2)^2+\nonumber \\
&&R(y_8)(\dot{q}_0-\dot{q}_3)^2+R(y_9)(\dot{q}_3-\dot{q}_4)^2+R(y_{10})\dot{q}_4^2+R(y_{11})(\dot{q}_0-\dot{q}_3)^2+R(y_{12})(\dot{q}_0-\dot{q}_4)^2\},
\end{eqnarray}
\end{widetext}
where the $\lambda_i$'s are the Lagrange multipliers.
These two functions can be easily generalized for the case of a
$N\times M$ network of different memristors, greatly facilitating
the attainment of the EOMs, the solution of which can be used to
solve, e.g., optimization problems such as mazes in a massively parallel way~\cite{pershin11d}.

\section{Hamilton formalism}\label{SEC:hf}
The counterpart of the Lagrange formalism is the Hamilton one, which is also generally the starting point for quantization. For non-dissipative systems, one can easily transform the Lagrangian to the Hamiltonian. In the
presence of dissipation instead, this task requires particular care.

Dissipation is the result of the tracing out of certain degrees of freedom resulting in an effective (reduced) description of the system of
interest in interaction with these degrees of freedom. However, the microscopic procedure of tracing out these degrees of freedom is most of the time difficult to carry out exactly, and dissipation is then introduced with
physically plausible "ad hoc" strategies.

There are several ways to add dissipation at the level of circuit Hamiltonians which range from complex Lagrangians (resulting in complex Hamiltonians) \cite{Dekker1975} to the addition of linear dissipative elements modeled by an infinite network of capacitors and inductors (see, e.g., \cite{Caldeira1983}). Since the discussion of dissipation in Hamiltonian dynamics would require an extensive
treatment by itself, here we limit our analysis to non-dissipative systems, and (except for the work-energy theorem discussed below) leave the Hamiltonian formalism of dissipative memory elements for a future publication.

\subsection{Canonically conjugate momenta}
Consider a non-dissipative network of memory elements. In order to write the Hamiltonian, we need to determine the momenta $p_j$ canonically conjugate to the variables $q_j$. These momenta are defined for circuits of voltage-controlled elements by
\begin{equation}
p_j\equiv \frac{\partial \mathcal{L}}{\partial \dot{q}_j}\label{ccmomenta},
\end{equation}
with the same definition for circuits of current-controlled elements except for $\dot{q}_j$ being replaced by $\dot{\phi}_j$. Looking at the expressions for the Lagrangians of the memory elements discussed above, one easily finds the physical meaning of $p_j$.
 In voltage-controlled circuits $p_j$ is the total of the fluxes generated \emph{by the inductors} in the $j$-th loop, while in current-controlled circuits, it is the charge in the $j$-th junction. In addition, if we define the canonically conjugate momentum to the internal degree of freedom $y_i$ as $z_i$, we readily find that for both voltage-controlled and current-controlled circuits
\begin{equation}
z_i\equiv \frac{\partial \mathcal{L}}{\partial
\dot{y}_i}=c_i\dot{y}_i\label{ccmomenta2}.
\end{equation}

The Hamilton's equations for voltage-controlled circuits then read
\begin{eqnarray}
\dot{q}_j=\frac{\partial H}{\partial p_j}\label{hamiltoneq1},\\
\dot{p}_j=-\frac{\partial H}{\partial q_j}\label{hamiltoneq2},
\end{eqnarray}
with $q_j$ replaced by $\phi_j$ for current-controlled circuits.

Using the results for the canonically conjugate momenta, we see that for voltage-controlled circuits Eq. (\ref{hamiltoneq1}) gives the current in the $j$-th loop in terms of the magnetic flux in the inductors in each loop, while Eq. (\ref{hamiltoneq2}) gives the change in the magnetic flux in the inductors in the $j$-th loop in terms of the charges in the loops. For current-controlled
circuits, on the other hand, Eq. (\ref{hamiltoneq1}) gives the potential in the $j$-th junction in terms of the charges in the circuit junctions, and Eq. (\ref{hamiltoneq2}) gives the current flowing into this junction in terms of the fluxes in the circuit junctions. The EOMs obtained here - while representing the same physics - are distinctly different from the ones in the Lagrangian formalism and therein lies their value.

With these results in mind, the Hamiltonian is defined as the Legendre transformation of the Lagrangian, namely
\begin{equation}
H=\sum\limits_j p_j \dot{q}_j + \sum\limits_i z_i \dot{y}_i -
\mathcal{L}\label{defhamiltonian}
\end{equation}
for voltage-controlled circuits, and with $q_j$ substituted by $\phi_j$ for current-controlled circuits. Since the kinetic energy in both cases is quadratic in $\dot{q}_j$ ($\dot{\phi}_j$ for current-controlled circuits), it is easy to see that Eq. (\ref{defhamiltonian}) reduces to $H=T+U$.

\subsection{Work-energy theorem}\label{SEC:hfWETh}

If we consider an arbitrary circuit with a number of voltage sources $V_{kj}$ (of the $k$-th voltage source in the $j$-th loop), we can define the work done by these sources, at any given time in an interval of time $dt$, on infinitesimal charges $dq_j$ in each of the indivisible loops. The total work done by all sources (which is not an exact differential) is then

\begin{equation}
\delta W=\sum_{k,j} V_{kj} dq_j\,.
\end{equation}

For non-dissipative circuits all this work goes into the variation of the internal energy $dE$ which can be computed from $T+U$ by subtracting the contribution from the voltage sources. The work-energy theorem in this case thus reads
\begin{equation}
\delta W=dE \label{w-end}\,.
\end{equation}

In the presence of current sources the work done is
\begin{equation}
\delta W=\sum_{k,j} I_{kj} (d\phi_k-d\phi_j)\,,
\end{equation}
where $I_{kj}$ is the current of the source between the $k$-th and $j$-th junctions (or 0 if none such source exists), and $d\phi_k-d\phi_j$ is the difference in flux on the two sides of the source. This work accounts for the change $dE$ of internal energy which derives from $T+U$ by subtracting the contribution from the current sources to give a balance formally equal to Eq.~(\ref{w-end}).

\subsection{Generalized Joule's first law}\label{SEC:hfWEThdiss}

In the dissipative case on the other hand we need to take into account that part of the work done by the voltage sources that goes into a "generalized heat" which accounts for the heat generated in the resistances (if present)
and the "heat" generated from the dissipative components of the state variables.

Mathematically, this amounts to
\begin{equation}
\delta W -dE=\sum\limits_j \frac{\partial \mathcal{H}}{\partial \dot{q}_j}\dot{q}_j
dt+\sum\limits_i
\frac{\partial \mathcal{H}}{\partial \dot{y}_i}\dot{y}_i dt\label{workenergy1}
\end{equation}
for voltage-controlled circuits. On the other hand, in the presence of current sources, Eq.~(\ref{workenergy1}) needs to be changed into
\begin{equation}
\delta W- dE=\sum\limits_j \frac{\partial \mathcal{H}}{\partial
\dot{\phi}_j}\dot{\phi}_j dt+\sum\limits_i
\frac{\partial \mathcal{H}}{\partial \dot{y}_i}\dot{y}_i dt\label{workenergy2}.
\end{equation}
Eqs. (\ref{workenergy1}) and
(\ref{workenergy2}) are the {\it generalized Joule's first laws} for dissipative systems
and are important yardsticks, together with the Euler-Lagrange
EOMs, to test the validity of a given Lagrangian formulation. We note in particular, the identification of the energy loss due to memory, given by the
last terms on the right-hand side of Eqs. (\ref{workenergy1}) and (\ref{workenergy2}), which are not present in the formulation of standard circuit elements.

\section{Quantization} \label{section5}
Having shown the Hamiltonian formulation for classical
non-dissipative circuits, we now embark on the quantization of
these Hamiltonians, which will be of importance at low temperatures and
mesoscopic/nanoscopic length scales. Instead of proceeding with a general circuit, in this case we find it more instructive to first work
out explicit examples. We consider first a
voltage source connected in series with a memcapacitive system and
a meminductive system. Then we look at a current source connected
in parallel with these systems. These two circuits can be realized
experimentally (see e.g., Ref.~\onlinecite{Johansson06a}) and are therefore ideal
test-beds for the concept of {\it memory quanta}, namely quantized
excitations of the memory degrees of freedom of these circuits.

\subsection{Example: series LC circuit}
\begin{figure}[t]
 \begin{center}
    \centerline{
    \mbox{(a)}
    \mbox{\includegraphics[width=3.60cm]{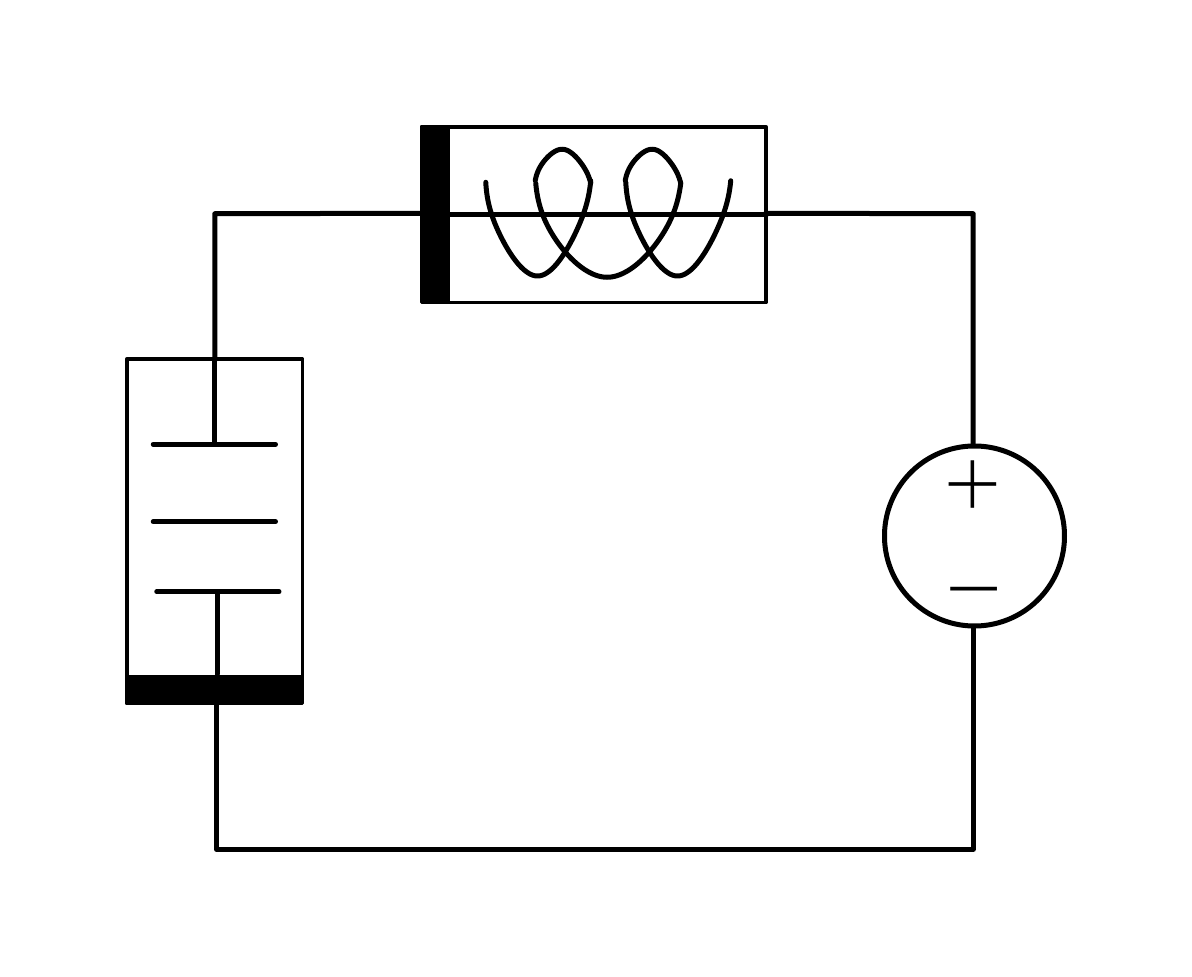}}
    \mbox{(b)}
    \mbox{\includegraphics[width=3.60cm]{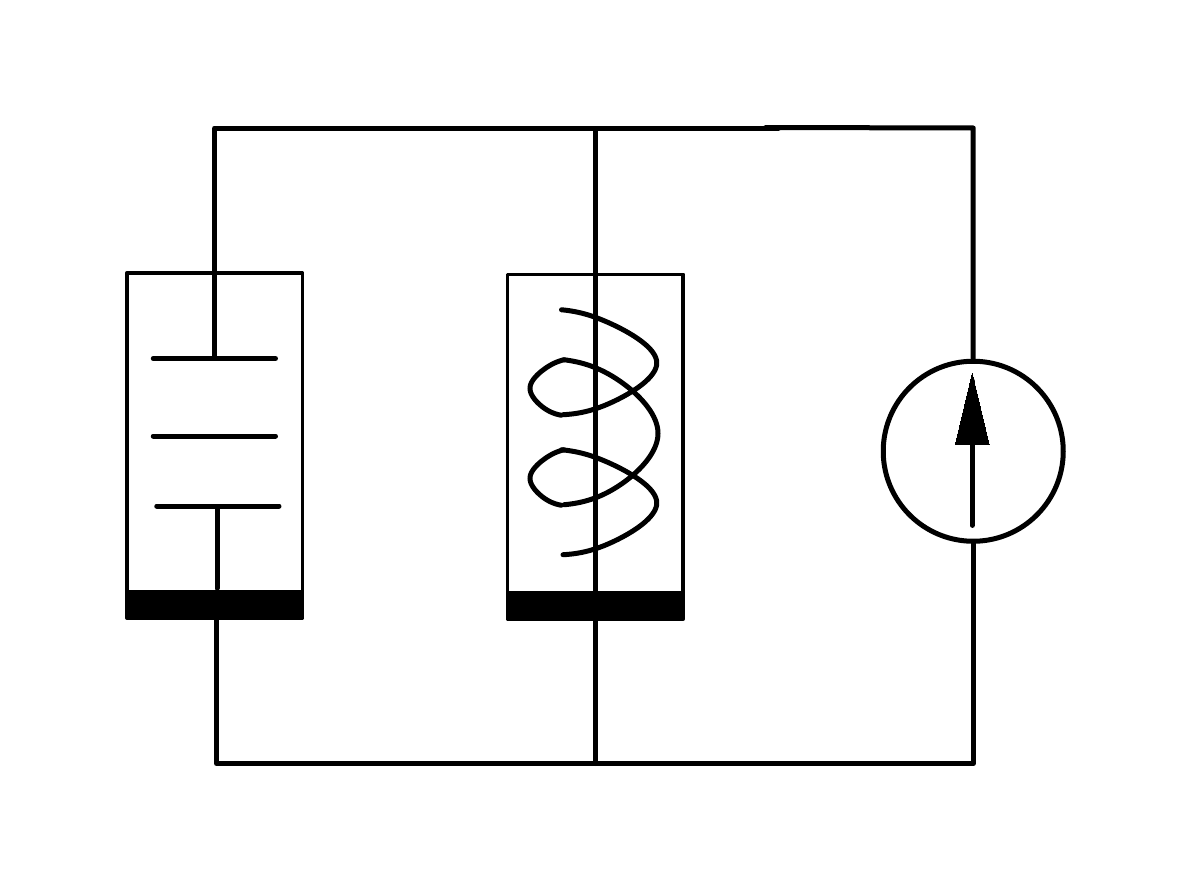}}
  }
\caption{\label{fig7}  (a) Schematic of a series
voltage-controlled memory element LC circuit. (b) Schematic of a
parallel current-controlled memory element LC circuit.}
 \end{center}
\end{figure}

We consider a voltage source connected in series with a
memcapacitive system and a meminductive one as depicted in Fig.
\ref{fig7}(a). The case of such a circuit with no memory
was quantized in previous works\cite{Devoret1997,zhang98a} and
will be generalized here. The Hamiltonian for this circuit is
found using Eqs. (\ref{eq:Tgen}) and (\ref{Cenergy}) for the
memcapacitive system and Eqs. (\ref{eq:TLP}) and (\ref{eq:Ugen}) for the
meminductive one, and reads
\begin{widetext}
\begin{equation}
H=T+U=\half\sum\limits_i c_{1i}^{-1}z_{1i}^2+\half\sum\limits_i
c_{2i}^{-1}z_{2i}^2+\frac{q^2}{2C(y_1,V_1,t)}+\frac{\phi^2}{2L(y_2,\phi,t)}+\widetilde{U}_1(y_1,V_1,t)+\widetilde{U}_2(y_2,\phi,t)-qV(t),
\label{series-LC-01}
\end{equation}
\end{widetext}
where the index 1 corresponds to the capacitor and the index 2 to
the inductor. $y_{ji}$ is the $i$-th memory coordinate of
the $j$-th memory element, and $z_{ji}$ is its canonically
conjugate momentum as defined in Eq. (\ref{ccmomenta2}). $q$ is the
charge flowing through the circuit, and $\phi$ is the flux on
the inductor. $V_1$ is the voltage on the capacitor and $V(t)$ is
the voltage of the source.

Under reasonable assumption of stability of the values of $y_1$
and $y_2$, we expand $\widetilde{U}_1(y_1,V_1,t)$ and
$\widetilde{U}_2(y_2,\phi_2,t)$ at their minima with respect to
$y_1$ and $y_2$, respectively. Several such minima can exist for
$y_1$ or $y_2$ in certain memory elements~\cite{pershin10a}. In
this case we should choose one minimum based on the initial
conditions. The definitions of $y_i$ are shifted by constants to
make them zero at their respective minima. This shift does not
affect the form of the other terms in the Hamiltonian. In
addition, we define $\Delta(C^{-1})(y_1,V_1,t)$ and
$\Delta(L^{-1})(y_2,\phi,t)$ by
\begin{eqnarray}
\Delta(C^{-1})(y_1,V_1,t)&\equiv& \frac{1}{C(y_1,V_1,t)}-\frac{1}{C_0(V_1,t)},\\
\Delta(L^{-1})(y_2,\phi,t)&\equiv&
\frac{1}{L(y_2,\phi,t)}-\frac{1}{L_0(\phi,t)},
\end{eqnarray}
where for brevity we have defined $C_0(V_1,t)\equiv C(0,V_1,t)$ and
$L_0(p,t)\equiv L(0,p,t)$.

Discarding constant terms in $\widetilde{U}_i$ and neglecting
higher order terms in $x_i$, we can write $\widetilde{U}_i
\approx \sum_j d_{ji}y_{ji}^2/2$, and the Hamiltonian takes the
form
\begin{eqnarray}
H&=&H_q+H_x+H_{int}\label{series-LC-02},\\
H_q&=&\frac{\phi^2}{2L_0}+\frac{q^2}{2C_0}-qV(t),\\
H_y&=&\half\sum\limits_{ij} c_{ji}^{-1}z_{ji}^2+\half
\sum\limits_{ij} d_{ji} y_{ji}^2,\\
H_{int}&=&\frac{\Delta (L^{-1}) \phi^2}{2}+\frac{\Delta (C^{-1})
q^2}{2},
\end{eqnarray}
where the Hamiltonian was divided into a "charge" part, $H_q$,
"memory" part, $H_y$, and the "interaction" part, $H_{int}$.

We next introduce the bosonic creation and annihilation operators
defined by
\begin{eqnarray}
a&=&\sqrt{\frac{L_0\omega_0}{2\hbar}}(q+\frac{i\phi}{L_0\omega_0}) \label{series-LC-04}\\
a^\dagger&=&\sqrt{\frac{L_0\omega_0}{2\hbar}}(q-\frac{i\phi}{L_0\omega_0}) \label{series-LC-05}\\
b_{ji}&=&\sqrt{\frac{c_{ji}\omega_{ji}}{2\hbar}}(y_{ji}+\frac{iz_{ji}}{c_{ji}\omega_{ji}}) \label{series-LC-06}\\
b_{ji}^\dagger&=&\sqrt{\frac{c_{ji}\omega_{ji}}{2\hbar}}(y_{ji}-\frac{iz_{ji}}{c_{ji}\omega_{ji}})\label{series-LC-07},
\end{eqnarray}
where $a^\dagger$ and $a$ ($b_{ji}^\dagger$ and $b_{ji}$) create
and destroy charge (memory) quanta, respectively.

The frequency of the charge
oscillator, $\omega_0$, is the circuit resonance frequency,
$(L_0C_0)^{-1/2}$, while the frequencies of the memory quanta
oscillators are analogously given by
\begin{equation}
\omega_{ji}\equiv \sqrt{\frac{d_{ji}}{c_{ji}}}.
\end{equation}
Plugging these relations into the Hamiltonian in Eq.
(\ref{series-LC-02}) finally gives the quantized form
\begin{eqnarray}
H_q&=&\hbar\omega_0(a^\dagger a + \half)-\sqrt{\frac{\hbar}{2L_0\omega_0}}(a+a^\dagger)V(t),\\
H_x&=&\sum\limits_{ij} \hbar\omega_{ji}(b_{ji}^\dagger b_{ji} + \half),\\
H_{int}&=&-\frac{\hbar L_0\omega_0}{4} \Delta
(L^{-1})(\widehat{y}_2,\widehat{\phi},t)
(a-a^\dagger)^2 \nonumber \\
&+&\frac{\hbar}{4L_0\omega_0} \Delta
(C^{-1})(\widehat{y}_1,\widehat{V}_1,t)(a+a^\dagger)^2\label{series-LC-03},
\end{eqnarray}
where $V_1$ is quantized by solving the equation
$V_1=q/C(y_1,V_1,t)$ to give $V_1=V_1(y_1,q,t)$ which translates
to $\widehat{V}_1=V_1(\widehat{y}_1,\widehat{q},t)$ after
quantization.

It is now clearly seen that $H_q$ includes only terms corresponding to the
charge quanta, while $H_y$ includes those corresponding to memory. $H_{int}$ couples the two quanta with a coupling
term that has at least three ladder operators.

\begin{figure}[t]
 \begin{center}
 \includegraphics[width=6cm]{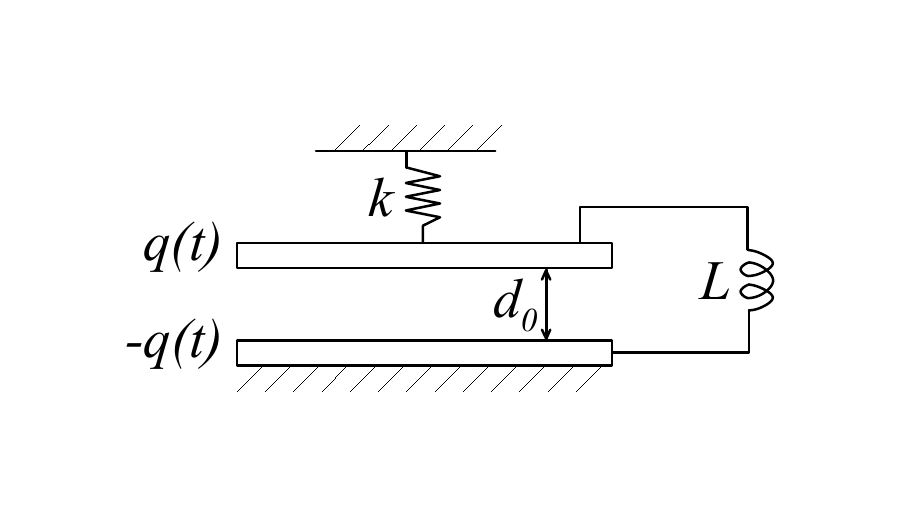}
\caption{\label{fig8}  Elastic memcapacitive system
connected to a normal inductor.}
 \end{center}
\end{figure}

A simple
example~\cite{pershin11a} that illustrates this result is a circuit
involving a normal inductor of inductance $L$ in series with a
memcapacitor that has its upper plate of mass $m$ hanging on a
spring with spring constant $k$ (Fig. \ref{fig8}). This memcapacitor could be a representation of, e.g., a
nano-electromechanical system~\cite{rfmemsbook,rfmemsbook1,evoy04a}. If the
displacement of the upper plate from its equilibrium position is
denoted by $y$ and its distance from the lower plate at this
position is $d_0$, the capacitance can be easily seen to be given
by~\cite{pershin11a}
\begin{equation}
C(x)=\frac{C_0}{1+y/d_0},
\end{equation}
where $C_0$ is the capacitance at equilibrium. Using the above
formalism to quantize the Hamiltonian and keeping only energy
conserving terms, we find Eq. (\ref{series-LC-03}) is reduced to
\begin{equation}
H_{int}=\frac{\hbar^{3/2}}{4\sqrt{2}}d_0^{-1}(LC_0)^{-1/2}(mk)^{-1/4}({a^\dagger}^2
b+b^\dagger a^2).
\end{equation}
This type of interaction is of the same kind as the one
encountered in the quantum treatment of second harmonic generation
in optics\cite{mandel95a}.

If the circuit can be built to satisfy
$(k/m)^{1/2}=2(LC_0)^{-1/2}$ the interaction will produce a
splitting of the degeneracy of the levels that is of {\it first order}
in $H_{int}$ and which may be large enough to be detected
experimentally. (See also the Conclusions for an order of magnitude estimate of when to expect quantum
effects to dominate.)

\subsection{Example: parallel LC circuit}
As a second example let us consider a parallel memory LC circuit as
plotted in Fig. \ref{fig7}(b). In this circuit a current source is
connected in parallel with a memcapacitive system and a
meminductive system. The Hamiltonian for this system can be found
utilizing Eqs. (\ref{CCM1}) and (\ref{CCM2}) for the former, and
Eqs. (\ref{CCL1}) and (\ref{CCL2}) for the latter resulting in
\begin{widetext}
\begin{equation}
H=T+U=\half\sum\limits_i c_{1i}^{-1}z_{1i}^2+\half\sum\limits_i
c_{2i}^{-1}z_{2i}^2+\frac{q^2}{2C(y_1,\phi,t)}+\frac{\phi^2}{2L(y_2,I_2,t)}+\widetilde{U}_1(y_1,C^{-1}\phi,t)+\widetilde{U}_2(y_2,LI_2,t)-\phi
I(t), \label{parallel-LC-01}
\end{equation}
\end{widetext}
with the same definitions as in Eq. (\ref{series-LC-01}), except
for $\phi$ being the flux in the inductor and $q$ being
the charge on the capacitor. $I_2$ is the current through the
inductor and $I(t)$ is the current of the source.

Proceeding in a completely analogous way to the previous
subsection with the definitions of the ladder operators in Eqs.
(\ref{series-LC-04})-(\ref{series-LC-07}) modified by the
substitutions $L_0\rightarrow C_0$ and $q\rightarrow \phi$, we
find the quantized Hamiltonian of this system to be
\begin{eqnarray}
H&=&H_q+H_y+H_{int}\\
H_q&=&\hbar\omega_0(a^\dagger a + \half)-\sqrt{\frac{\hbar}{2C_0\omega_0}}(a+a^\dagger)I(t),\\
H_y&=&\sum\limits_{ij} \hbar\omega_{ji}(b_{ji}^\dagger b_{ji} + \half),\\
H_{int}&=&-\frac{\hbar C_0\omega_0}{4} \Delta
(L^{-1})(\widehat{x}_2,\widehat{I}_2,t)
(a-a^\dagger)^2 \nonumber \\
&+&\frac{\hbar}{4C_0\omega_0} \Delta
(C^{-1})(\widehat{x}_1,\widehat{q},t)(a+a^\dagger)^2\label{parallel-LC-02},
\end{eqnarray}
where $I_2$ is quantized by solving the equation
$I_2=\phi/L(y_2,I_2,t)$ to give $I_2=I_2(y_2,\phi,t)$ which
reduces to $\widehat{I}_2=I_2(\widehat{y}_2,\widehat{\phi},t)$
after quantization. This Hamiltonian is very similar to the one
obtained for the series LC circuit. We will now show these two
Hamiltonians to be the basic building blocks for the quantized
Hamiltonian of a general circuit with non-dissipative elements.

\subsection{General circuit}
We now proceed to find the quantized Hamiltonian for a general
circuit network of memcapacitive systems, meminductive systems and
voltage sources. Such a circuit can be divided into indivisible
loops each with charge $q_k$ as noted in Sec.
\ref{SEC:lagrangian-general-circuit}. Using the methods of that
section to find $T$ and $U$, one can write the Hamiltonian for the
network as $H=T+U$, which, after the substitution of $q_k$ and
$\phi_k$ with ladder operators using Eqs.
(\ref{series-LC-04})-(\ref{series-LC-07}), reduces, apart from the
interaction part, to a bilinear combination of them which is known
to be exactly diagonalizable by, e.g., a linear canonical
transformation.

With regards to the quantization of the Hamiltonian of
current-controlled circuits, the process is similar. We denote the
flux in each junction of the network with $\phi_k$ with a
corresponding $q_k$ being the charge in the junction. Using the
methods of Sec. \ref{SEC:lagrangian-general-circuit}, we write the
Hamiltonian $H=T+V$ and then quantize it by writing $\phi_k$ and
$p_k$ in terms of ladder operators using the transformation from
the previous subsection. Like in the voltage-controlled case, the
non-interacting part of the Hamiltonian is again bilinear and can
be diagonalized.

\section{Conclusions} \label{section6}
To summarize, in this work we introduced the general Lagrangian
formulation for the three basic memory elements: memristive,
memcapacitive and meminductive systems and defined a fourth memory
element, a \emph{mutual meminductive system}, for which we also
gave the Lagrange formalism. We showed how to write the Lagrangian
for a general circuit, including one with current sources. The
examples given for the Lagrangian formalism demonstrated that
writing the Lagrangian and dissipation potential should be the
preferred choice for finding the EOMs of large memory element
networks.

The Hamiltonian formalism for electric circuits was also
generalized to include memory, although only for non-dissipative
elements. As in previous works~\cite{Devoret1997}, we have found that the
canonically conjugate momentum of charge is the flux and
vice versa. The Generalized Joule's first law was given for
general circuits including ones with memory elements. This law can
be used to verify the correctness of a given Lagrangian
formulation. Lastly, we presented a scheme for the quantization of
a general non-dissipative memory element circuit.

The quantum treatment of memory elements, and in particular the
example given in the text of a memcapacitor in series with an
inductor (Fig. \ref{fig8}), begs the question of under which
conditions one can measure quantum effects in these systems. For
quantum effects to be easily measurable, both the thermal
fluctuation energy and the width of the energy levels should be
smaller than the oscillator energy quantum~\cite{Devoret1997},
i.e., $k_BT\ll \hbar \omega_0$ and $\mathcal{Q}\gg 1$, where
$\mathcal{Q}=\omega_0 R/L$ is the quality factor of the
oscillator, $R$ the loop resistance, and $L$ is the inductance.
Possible values for the capacitance and inductance in mesoscopic
circuits can be taken to be $10^{-15}$ F \cite{Watanabe03a} and
$10^{-10}$ H \cite{SQUID_book}, respectively. If one assumes a
temperature of $T=20$ mK and circuit resistance of 10 $\Omega$ or
less, both conditions mentioned above are satisfied. As noted for
the example above, the degeneracy condition, satisfiable by a
memcapacitor~\cite{partensky2002-1}, will lead to an
experimentally detectable splitting of the degenerate energy
levels as a result of the interaction between the memory quanta
and charge quanta.

Future research in this field may include extending the Hamiltonian
formalism to dissipative circuits. One way to do this is, e.g., via a path-integral formulation~\cite{Path_int_book} of memory elements. Along a parallel line, we expect the
Lagrangian formalism discussed here to be of great value in the analysis of complex networks with memory, which offer both fundamental and applied research opportunities.

\section*{Acknowledgments}
This work has been
partially funded by the NSF grant No. DMR-0802830. One of us (MD) is grateful to the Scuola Normale Superiore of Pisa for the hospitality during a visit where part of this
work has been written, and to S. Pugnetti and R. Fazio for useful discussions.

\section*{References}
%
\bibliography{memcapacitor}

\begin{thebibliography}{10}%
\makeatletter
\providecommand \@ifxundefined [1]{%
 \ifx #1\undefined \expandafter \@firstoftwo
 \else \expandafter \@secondoftwo
\fi
}%
\providecommand \@ifnum [1]{%
 \ifnum #1\expandafter \@firstoftwo
 \else \expandafter \@secondoftwo
\fi
}%
\providecommand \enquote [1]{``#1''}%
\providecommand \bibnamefont  [1]{#1}%
\providecommand \bibfnamefont [1]{#1}%
\providecommand \citenamefont [1]{#1}%
\providecommand\href[0]{\@sanitize\@href}%
\providecommand\@href[1]{\endgroup\@@startlink{#1}\endgroup\@@href}%
\providecommand\@@href[1]{#1\@@endlink}%
\providecommand \@sanitize [0]{\begingroup\catcode`\&12\catcode`\#12\relax}%
\@ifxundefined \pdfoutput {\@firstoftwo}{%
 \@ifnum{\z@=\pdfoutput}{\@firstoftwo}{\@secondoftwo}%
}{%
 \providecommand\@@startlink[1]{\leavevmode\special{html:<a href="#1">}}%
 \providecommand\@@endlink[0]{\special{html:</a>}}%
}{%
 \providecommand\@@startlink[1]{%
  \leavevmode
  \pdfstartlink
   attr{/Border[0 0 1 ]/H/I/C[0 1 1]}%
   user{/Subtype/Link/A<</Type/Action/S/URI/URI(#1)>>}%
  \relax
 }%
 \providecommand\@@endlink[0]{\pdfendlink}%
}%
\providecommand \url  [0]{\begingroup\@sanitize \@url }%
\providecommand \@url [1]{\endgroup\@href {#1}{\urlprefix}}%
\providecommand \urlprefix [0]{URL }%
\providecommand \Eprint[0]{\href }%
\@ifxundefined \urlstyle {%
  \providecommand \doi [1]{doi:\discretionary{}{}{}#1}%
}{%
  \providecommand \doi [0]{doi:\discretionary{}{}{}\begingroup
  \urlstyle{rm}\Url }%
}%
\providecommand \doibase [0]{http://dx.doi.org/}%
\providecommand \Doi[1]{\href{\doibase#1}}%
\providecommand \bibAnnote [3]{%
  \BibitemShut{#1}%
  \begin{quotation}\noindent
    \textsc{Key:}\ #2\\\textsc{Annotation:}\ #3%
  \end{quotation}%
}%
\providecommand \bibAnnoteFile [2]{%
  \IfFileExists{#2}{\bibAnnote {#1} {#2} {\input{#2}}}{}%
}%
\providecommand \typeout [0]{\immediate \write \m@ne }%
\providecommand \selectlanguage [0]{\@gobble}%
\providecommand \bibinfo [0]{\@secondoftwo}%
\providecommand \bibfield [0]{\@secondoftwo}%
\providecommand \translation [1]{[#1]}%
\providecommand \BibitemOpen[0]{}%
\providecommand \bibitemStop [0]{}%
\providecommand \bibitemNoStop [0]{.\EOS\space}%
\providecommand \EOS [0]{\spacefactor3000\relax}%
\providecommand \BibitemShut [1]{\csname bibitem#1\endcsname}%
\bibitem{chua71a}%
  \BibitemOpen
  \bibfield{author}{%
  \bibinfo {author} {\bibfnamefont{L.~O.}\ \bibnamefont{Chua}},\ }%
  \bibfield{journal}{%
  \bibinfo {journal} {{IEEE} Trans. Circuit Theory}\ }%
  \textbf{\bibinfo {volume} {18}},\ \bibinfo {pages} {507} (\bibinfo {year}
  {1971})%
  \bibAnnoteFile{NoStop}{chua71a}%
\bibitem{chua76a}%
  \BibitemOpen
  \bibfield{author}{%
  \bibinfo {author} {\bibfnamefont{L.~O.}\ \bibnamefont{Chua}}\ and\ \bibinfo
  {author} {\bibfnamefont{S.~M.}\ \bibnamefont{Kang}},\ }%
  \bibfield{journal}{%
  \bibinfo {journal} {Proc. {IEEE}}\ }%
  \textbf{\bibinfo {volume} {64}},\ \bibinfo {pages} {209} (\bibinfo {year}
  {1976})%
  \bibAnnoteFile{NoStop}{chua76a}%
\bibitem{diventra09a}%
  \BibitemOpen
  \bibfield{author}{%
  \bibinfo {author} {\bibfnamefont{M.}~\bibnamefont{{Di Ventra}}}, \bibinfo
  {author} {\bibfnamefont{Y.~V.}\ \bibnamefont{Pershin}},\ and\ \bibinfo
  {author} {\bibfnamefont{L.~O.}\ \bibnamefont{Chua}},\ }%
  \bibfield{journal}{%
  \Doi{10.1109/JPROC.2009.2021077}{\bibinfo {journal} {Proc. {IEEE}}}\ }%
  \textbf{\bibinfo {volume} {97}},\ \bibinfo {pages} {1717} (\bibinfo {year}
  {2009})%
  \bibAnnoteFile{NoStop}{diventra09a}%
\bibitem{Green07a}%
  \BibitemOpen
  \bibfield{author}{%
  \bibinfo {author} {\bibfnamefont{J.~E.}\ \bibnamefont{Green}}, \bibinfo
  {author} {\bibfnamefont{J.~W.}\ \bibnamefont{Choi}}, \bibinfo {author}
  {\bibfnamefont{A.}~\bibnamefont{Boukai}}, \bibinfo {author}
  {\bibfnamefont{Y.}~\bibnamefont{Bunimovich}}, \bibinfo {author}
  {\bibfnamefont{E.}~\bibnamefont{Johnston-Halperin}}, \bibinfo {author}
  {\bibfnamefont{E.}~\bibnamefont{DeIonno}}, \bibinfo {author}
  {\bibfnamefont{Y.}~\bibnamefont{Luo}}, \bibinfo {author}
  {\bibfnamefont{B.~A.}\ \bibnamefont{Sheriff}}, \bibinfo {author}
  {\bibfnamefont{K.}~\bibnamefont{Xu}}, \bibinfo {author}
  {\bibfnamefont{Y.~S.}\ \bibnamefont{Shin}}, \bibinfo {author}
  {\bibfnamefont{H.-R.}\ \bibnamefont{Tseng}}, \bibinfo {author}
  {\bibfnamefont{J.~F.}\ \bibnamefont{Stoddart}},\ and\ \bibinfo {author}
  {\bibfnamefont{J.~R.}\ \bibnamefont{Heath}},\ }%
  \bibfield{journal}{%
  \bibinfo {journal} {Nature}\ }%
  \textbf{\bibinfo {volume} {445}},\ \bibinfo {pages} {414} (\bibinfo {year}
  {2007})%
  \bibAnnoteFile{NoStop}{Green07a}%
\bibitem{Karg08a}%
  \BibitemOpen
  \bibfield{author}{%
  \bibinfo {author} {\bibfnamefont{S.~F.}\ \bibnamefont{Karg}}, \bibinfo
  {author} {\bibfnamefont{G.~I.}\ \bibnamefont{Meijer}}, \bibinfo {author}
  {\bibfnamefont{J.~G.}\ \bibnamefont{Bednorz}}, \bibinfo {author}
  {\bibfnamefont{C.~T.}\ \bibnamefont{Rettner}}, \bibinfo {author}
  {\bibfnamefont{A.~G.}\ \bibnamefont{Schrott}}, \bibinfo {author}
  {\bibfnamefont{E.~A.}\ \bibnamefont{Joseph}}, \bibinfo {author}
  {\bibfnamefont{C.~H.}\ \bibnamefont{Lam}}, \bibinfo {author}
  {\bibfnamefont{M.}~\bibnamefont{Janousch}}, \bibinfo {author}
  {\bibfnamefont{U.}~\bibnamefont{Staub}}, \bibinfo {author}
  {\bibfnamefont{F.}~\bibnamefont{La~Mattina}}, \bibinfo {author}
  {\bibfnamefont{S.~F.}\ \bibnamefont{Alvarado}}, \bibinfo {author}
  {\bibfnamefont{D.}~\bibnamefont{Widmer}}, \bibinfo {author}
  {\bibfnamefont{R.}~\bibnamefont{Stutz}}, \bibinfo {author}
  {\bibfnamefont{U.}~\bibnamefont{Drechsler}},\ and\ \bibinfo {author}
  {\bibfnamefont{D.}~\bibnamefont{Caimi}},\ }%
  \bibfield{journal}{%
  \bibinfo {journal} {IBM J. Res. Dev.}\ }%
  \textbf{\bibinfo {volume} {52}},\ \bibinfo {pages} {481} (\bibinfo {month}
  {JUL-SEP}\ \bibinfo {year} {2008})%
  \bibAnnoteFile{NoStop}{Karg08a}%
\bibitem{Sawa08a}%
  \BibitemOpen
  \bibfield{author}{%
  \bibinfo {author} {\bibfnamefont{A.}~\bibnamefont{Sawa}},\ }%
  \bibfield{journal}{%
  \bibinfo {journal} {Mat. Today}\ }%
  \textbf{\bibinfo {volume} {11}},\ \bibinfo {pages} {28} (\bibinfo {year}
  {2008})%
  \bibAnnoteFile{NoStop}{Sawa08a}%
\bibitem{pershin09c}%
  \BibitemOpen
  \bibfield{author}{%
  \bibinfo {author} {\bibfnamefont{Y.~V.}\ \bibnamefont{Pershin}}\ and\
  \bibinfo {author} {\bibfnamefont{M.}~\bibnamefont{{Di Ventra}}},\ }%
  \bibfield{journal}{%
  \bibinfo {journal} {Neural {N}etworks}\ }%
  \textbf{\bibinfo {volume} {23}},\ \bibinfo {pages} {881} (\bibinfo {year}
  {2010})%
  \bibAnnoteFile{NoStop}{pershin09c}%
\bibitem{jo10a}%
  \BibitemOpen
  \bibfield{author}{%
  \bibinfo {author} {\bibfnamefont{S.~H.}\ \bibnamefont{Jo}}, \bibinfo {author}
  {\bibfnamefont{T.}~\bibnamefont{Chang}}, \bibinfo {author}
  {\bibfnamefont{I.}~\bibnamefont{Ebong}}, \bibinfo {author}
  {\bibfnamefont{B.~B.}\ \bibnamefont{Bhadviya}}, \bibinfo {author}
  {\bibfnamefont{P.}~\bibnamefont{Mazumder}},\ and\ \bibinfo {author}
  {\bibfnamefont{W.}~\bibnamefont{Lu}},\ }%
  \bibfield{journal}{%
  \bibinfo {journal} {Nano Lett.}\ }%
  \textbf{\bibinfo {volume} {10}},\ \bibinfo {pages} {1297} (\bibinfo {year}
  {2010})%
  \bibAnnoteFile{NoStop}{jo10a}%
\bibitem{Choi09a}%
  \BibitemOpen
  \bibfield{author}{%
  \bibinfo {author} {\bibfnamefont{H.}~\bibnamefont{Choi}}, \bibinfo {author}
  {\bibfnamefont{H.}~\bibnamefont{Jung}}, \bibinfo {author}
  {\bibfnamefont{J.}~\bibnamefont{Lee}}, \bibinfo {author}
  {\bibfnamefont{J.}~\bibnamefont{Yoon}}, \bibinfo {author}
  {\bibfnamefont{J.}~\bibnamefont{Park}}, \bibinfo {author}
  {\bibfnamefont{D.-J.}\ \bibnamefont{Seong}}, \bibinfo {author}
  {\bibfnamefont{W.}~\bibnamefont{Lee}}, \bibinfo {author}
  {\bibfnamefont{M.}~\bibnamefont{Hasan}}, \bibinfo {author}
  {\bibfnamefont{G.-Y.}\ \bibnamefont{Jung}},\ and\ \bibinfo {author}
  {\bibfnamefont{H.}~\bibnamefont{Hwang}},\ }%
  \bibfield{journal}{%
  \bibinfo {journal} {Nanotechn.}\ }%
  \textbf{\bibinfo {volume} {20}},\ \bibinfo {pages} {345201} (\bibinfo {year}
  {2009})%
  \bibAnnoteFile{NoStop}{Choi09a}%
\bibitem{Lai10a}%
  \BibitemOpen
  \bibfield{author}{%
  \bibinfo {author} {\bibfnamefont{Q.}~\bibnamefont{Lai}}, \bibinfo {author}
  {\bibfnamefont{L.}~\bibnamefont{Zhang}}, \bibinfo {author}
  {\bibfnamefont{Z.}~\bibnamefont{Li}}, \bibinfo {author}
  {\bibfnamefont{W.~F.}\ \bibnamefont{Stickle}}, \bibinfo {author}
  {\bibfnamefont{R.~S.}\ \bibnamefont{Williams}},\ and\ \bibinfo {author}
  {\bibfnamefont{Y.}~\bibnamefont{Chen}},\ }%
  \bibfield{journal}{%
  \bibinfo {journal} {Adv. Mat.}\ }%
  \textbf{\bibinfo {volume} {22}},\ \bibinfo {pages} {2448} (\bibinfo {year}
  {2010})%
  \bibAnnoteFile{NoStop}{Lai10a}%
\bibitem{Alibart10a}%
  \BibitemOpen
  \bibfield{author}{%
  \bibinfo {author} {\bibfnamefont{F.}~\bibnamefont{Alibart}}, \bibinfo
  {author} {\bibfnamefont{S.}~\bibnamefont{Pleutin}}, \bibinfo {author}
  {\bibfnamefont{D.}~\bibnamefont{Guerin}}, \bibinfo {author}
  {\bibfnamefont{C.}~\bibnamefont{Novembre}}, \bibinfo {author}
  {\bibfnamefont{S.}~\bibnamefont{Lenfant}}, \bibinfo {author}
  {\bibfnamefont{K.}~\bibnamefont{Lmimouni}}, \bibinfo {author}
  {\bibfnamefont{C.}~\bibnamefont{Gamrat}},\ and\ \bibinfo {author}
  {\bibfnamefont{D.}~\bibnamefont{Vuillaume}},\ }%
  \bibfield{journal}{%
  \bibinfo {journal} {Adv. Funct. Mat.}\ }%
  \textbf{\bibinfo {volume} {20}},\ \bibinfo {pages} {330} (\bibinfo {year}
  {2010})%
  \bibAnnoteFile{NoStop}{Alibart10a}%
\bibitem{fontana10a}%
  \BibitemOpen
  \bibfield{author}{%
  \bibinfo {author} {\bibfnamefont{M.~P.}\ \bibnamefont{Fontana}}}%
   (\bibinfo {year} {2010}),\ \bibinfo {note} {private communication}%
  \bibAnnoteFile{NoStop}{fontana10a}%
\bibitem{pershin09b}%
  \BibitemOpen
  \bibfield{author}{%
  \bibinfo {author} {\bibfnamefont{Y.~V.}\ \bibnamefont{Pershin}}, \bibinfo
  {author} {\bibfnamefont{S.}~\bibnamefont{La~Fontaine}},\ and\ \bibinfo
  {author} {\bibfnamefont{M.}~\bibnamefont{{Di Ventra}}},\ }%
  \bibfield{journal}{%
  \bibinfo {journal} {Phys. Rev. E}\ }%
  \textbf{\bibinfo {volume} {80}},\ \bibinfo {pages} {021926} (\bibinfo {year}
  {2009})%
  \bibAnnoteFile{NoStop}{pershin09b}%
\bibitem{Johnsen11a}%
  \BibitemOpen
  \bibfield{author}{%
  \bibinfo {author} {\bibfnamefont{G.~K.}\ \bibnamefont{Johnsen}}, \bibinfo
  {author} {\bibfnamefont{C.~A.}\ \bibnamefont{L\"utken}}, \bibinfo {author}
  {\bibfnamefont{O.~G.}\ \bibnamefont{Martinsen}},\ and\ \bibinfo {author}
  {\bibfnamefont{S.}~\bibnamefont{Grimnes}},\ }%
  \bibfield{journal}{%
  \Doi{10.1103/PhysRevE.83.031916}{\bibinfo {journal} {Phys. Rev. E}}\ }%
  \textbf{\bibinfo {volume} {83}},\ \bibinfo {pages} {031916} (\bibinfo {month}
  {Mar}\ \bibinfo {year} {2011})%
  \bibAnnoteFile{NoStop}{Johnsen11a}%
\bibitem{pershin08a}%
  \BibitemOpen
  \bibfield{author}{%
  \bibinfo {author} {\bibfnamefont{Y.~V.}\ \bibnamefont{Pershin}}\ and\
  \bibinfo {author} {\bibfnamefont{M.}~\bibnamefont{{Di Ventra}}},\ }%
  \bibfield{journal}{%
  \bibinfo {journal} {Phys. Rev. B}\ }%
  \textbf{\bibinfo {volume} {78}},\ \bibinfo {pages} {113309} (\bibinfo {year}
  {2008})%
  \bibAnnoteFile{NoStop}{pershin08a}%
\bibitem{wang09a}%
  \BibitemOpen
  \bibfield{author}{%
  \bibinfo {author} {\bibfnamefont{X.}~\bibnamefont{Wang}}, \bibinfo {author}
  {\bibfnamefont{Y.}~\bibnamefont{Chen}}, \bibinfo {author}
  {\bibfnamefont{H.}~\bibnamefont{Xi}}, \bibinfo {author}
  {\bibfnamefont{H.}~\bibnamefont{Li}},\ and\ \bibinfo {author}
  {\bibfnamefont{D.}~\bibnamefont{Dimitrov}},\ }%
  \bibfield{journal}{%
  \bibinfo {journal} {El. Dev. Lett.}\ }%
  \textbf{\bibinfo {volume} {30}},\ \bibinfo {pages} {294} (\bibinfo {year}
  {2009})%
  \bibAnnoteFile{NoStop}{wang09a}%
\bibitem{strukov08a}%
  \BibitemOpen
  \bibfield{author}{%
  \bibinfo {author} {\bibfnamefont{D.~B.}\ \bibnamefont{Strukov}}, \bibinfo
  {author} {\bibfnamefont{G.~S.}\ \bibnamefont{Snider}}, \bibinfo {author}
  {\bibfnamefont{D.~R.}\ \bibnamefont{Stewart}},\ and\ \bibinfo {author}
  {\bibfnamefont{R.~S.}\ \bibnamefont{Williams}},\ }%
  \bibfield{journal}{%
  \bibinfo {journal} {Nature}\ }%
  \textbf{\bibinfo {volume} {453}},\ \bibinfo {pages} {80} (\bibinfo {year}
  {2008})%
  \bibAnnoteFile{NoStop}{strukov08a}%
\bibitem{pershin11a}%
  \BibitemOpen
  \bibfield{author}{%
  \bibinfo {author} {\bibfnamefont{Y.~V.}\ \bibnamefont{Pershin}}\ and\
  \bibinfo {author} {\bibfnamefont{M.}~\bibnamefont{Di~Ventra}},\ }%
  \bibfield{journal}{%
  \bibinfo {journal} {Advances in Physics}\ }%
  \textbf{\bibinfo {volume} {60}},\ \bibinfo {pages} {145} (\bibinfo {year}
  {2011})%
  \bibAnnoteFile{NoStop}{pershin11a}%
\bibitem{ourrecentMT}%
  \BibitemOpen
  \bibfield{author}{%
  \bibinfo {author} {\bibfnamefont{M.}~\bibnamefont{Di~Ventra}}\ and\ \bibinfo
  {author} {\bibfnamefont{Y.~V.}\ \bibnamefont{Pershin}},\ }%
  \bibfield{journal}{%
  \bibinfo {journal} {Materials Today}\ }%
  \textbf{\bibinfo {volume} {14}},\ \bibinfo {pages} {584} (\bibinfo {year}
  {2011})%
  \bibAnnoteFile{NoStop}{ourrecentMT}%
\bibitem{pershin10c}%
  \BibitemOpen
  \bibfield{author}{%
  \bibinfo {author} {\bibfnamefont{Y.~V.}\ \bibnamefont{Pershin}}\ and\
  \bibinfo {author} {\bibfnamefont{M.}~\bibnamefont{{Di Ventra}}},\ }%
  \bibfield{journal}{%
  \bibinfo {journal} {Proc. {IEEE} (in press); ar{X}ive:1009.6025}}%
   (\bibinfo {year} {2011})%
  \bibAnnoteFile{NoStop}{pershin10c}%
\bibitem{pershin11d}%
  \BibitemOpen
  \bibfield{author}{%
  \bibinfo {author} {\bibfnamefont{Y.~V.}\ \bibnamefont{Pershin}}\ and\
  \bibinfo {author} {\bibfnamefont{M.}~\bibnamefont{Di~Ventra}},\ }%
  \bibfield{journal}{%
  \bibinfo {journal} {Phys. Rev. E}\ }%
  \textbf{\bibinfo {volume} {84}},\ \bibinfo {pages} {046703} (\bibinfo {year}
  {2011})%
  \bibAnnoteFile{NoStop}{pershin11d}%
\bibitem{Driscoll10b}%
  \BibitemOpen
  \bibfield{author}{%
  \bibinfo {author} {\bibfnamefont{T.}~\bibnamefont{Driscoll}}, \bibinfo
  {author} {\bibfnamefont{Y.~V.}\ \bibnamefont{Pershin}}, \bibinfo {author}
  {\bibfnamefont{D.~N.}\ \bibnamefont{Basov}},\ and\ \bibinfo {author}
  {\bibfnamefont{M.}~\bibnamefont{{Di Ventra}}},\ }%
  \bibfield{journal}{%
  \bibinfo {journal} {Appl. Phys. A}\ }%
  \textbf{\bibinfo {volume} {102}},\ \bibinfo {pages} {885} (\bibinfo {year}
  {2011})%
  \bibAnnoteFile{NoStop}{Driscoll10b}%
\bibitem{diventra09b}%
  \BibitemOpen
  \bibfield{author}{%
  \bibinfo {author} {\bibfnamefont{M.}~\bibnamefont{{Di Ventra}}}, \bibinfo
  {author} {\bibfnamefont{Y.~V.}\ \bibnamefont{Pershin}},\ and\ \bibinfo
  {author} {\bibfnamefont{L.~O.}\ \bibnamefont{Chua}},\ }%
  \bibfield{journal}{%
  \bibinfo {journal} {Proc. {IEEE}}\ }%
  \textbf{\bibinfo {volume} {97}},\ \bibinfo {pages} {1371} (\bibinfo {year}
  {2009})%
  \bibAnnoteFile{NoStop}{diventra09b}%
\bibitem{Maxbook}%
  \BibitemOpen
  \bibfield{author}{%
  \bibinfo {author} {\bibfnamefont{M.}~\bibnamefont{Di~Ventra}},\ }%
  \emph{\bibinfo {title} {Electrical Transport in Nanoscale Systems}}\
  (\bibinfo {publisher} {Cambridge University Press},\ \bibinfo {year} {2008})%
  \bibAnnoteFile{NoStop}{Maxbook}%
\bibitem{Breuer2002a}%
  \BibitemOpen
  \bibfield{author}{%
  \bibinfo {author} {\bibfnamefont{H.}~\bibnamefont{Breuer}}\ and\ \bibinfo
  {author} {\bibfnamefont{F.}~\bibnamefont{Petruccione}},\ }%
  \emph{\bibinfo {title} {The Theory of Open Quantum Systems}}\ (\bibinfo
  {publisher} {Oxford University Press},\ \bibinfo {year} {2002})%
  \bibAnnoteFile{NoStop}{Breuer2002a}%
\bibitem{Devoret1997}%
  \BibitemOpen
  \bibfield{author}{%
  \bibinfo {author} {\bibfnamefont{M.}~\bibnamefont{Devoret}},\ }%
  in\ \emph{\bibinfo {booktitle} {Quantum Fluctuations (Les Houches Session
  LXIII)}},\ \bibinfo {editor} {edited by\ \bibinfo {editor}
  {\bibfnamefont{S.}~\bibnamefont{Reynaud}}, \bibinfo {editor}
  {\bibfnamefont{E.}~\bibnamefont{Giacobino}},\ and\ \bibinfo {editor}
  {\bibfnamefont{J.}~\bibnamefont{Zinn-Justin}}}\ (\bibinfo {publisher}
  {Elsevier, New York},\ \bibinfo {year} {1997})%
  \bibAnnoteFile{NoStop}{Devoret1997}%
\bibitem{shragowitz88a}%
  \BibitemOpen
  \bibfield{author}{%
  \bibinfo {author} {\bibfnamefont{E.}~\bibnamefont{Shragowitz}}\ and\ \bibinfo
  {author} {\bibfnamefont{E.}~\bibnamefont{Gerlovin}},\ }%
  \bibfield{journal}{%
  \bibinfo {journal} {Int. J. Circ. Theor. Appl.}\ }%
  \textbf{\bibinfo {volume} {16}},\ \bibinfo {pages} {129} (\bibinfo {year}
  {1988})%
  \bibAnnoteFile{NoStop}{shragowitz88a}%
\bibitem{jeltsema10a}%
  \BibitemOpen
  \bibfield{author}{%
  \bibinfo {author} {\bibfnamefont{D.}~\bibnamefont{Jeltsema}}\ and\ \bibinfo
  {author} {\bibfnamefont{A.~J.}\ \bibnamefont{{van der Schaf}}},\ }%
  \bibfield{journal}{%
  \bibinfo {journal} {Mathematical and Computer Modelling of Dynamical
  Systems}\ }%
  \textbf{\bibinfo {volume} {16}},\ \bibinfo {pages} {75} (\bibinfo {year}
  {2010})%
  \bibAnnoteFile{NoStop}{jeltsema10a}%
\bibitem{Helmholtz1853a}%
  \BibitemOpen
  \bibfield{author}{%
  \bibinfo {author} {\bibfnamefont{H.}~\bibnamefont{Helmholtz}},\ }%
  \bibfield{journal}{%
  \bibinfo {journal} {Annalen der Physik und Chemie}\ }%
  \textbf{\bibinfo {volume} {89}},\ \bibinfo {pages} {211} (\bibinfo {year}
  {1883})%
  \bibAnnoteFile{NoStop}{Helmholtz1853a}%
\bibitem{thevenin1883a}%
  \BibitemOpen
  \bibfield{author}{%
  \bibinfo {author} {\bibfnamefont{L.}~\bibnamefont{Th\'{e}venin}},\ }%
  \bibfield{journal}{%
  \bibinfo {journal} {Annales T\'{e}l\'{e}graphiques}\ }%
  \textbf{\bibinfo {volume} {10}},\ \bibinfo {pages} {222} (\bibinfo {year}
  {1883})%
  \bibAnnoteFile{NoStop}{thevenin1883a}%
\bibitem{goldstein01a}%
  \BibitemOpen
  \bibfield{author}{%
  \bibinfo {author} {\bibfnamefont{H.}~\bibnamefont{Goldstein}}, \bibinfo
  {author} {\bibfnamefont{C.~P.}\ \bibnamefont{Poole}},\ and\ \bibinfo {author}
  {\bibfnamefont{J.~L.}\ \bibnamefont{Safko}},\ }%
  \emph{\bibinfo {title} {{Classical Mechanics (3rd Edition)}}},\ \bibinfo
  {edition} {3rd}\ ed.\ (\bibinfo {publisher} {Addison Wesley},\ \bibinfo
  {year} {2001})%
  \bibAnnoteFile{NoStop}{goldstein01a}%
\bibitem{NortTheorem}%
  \BibitemOpen
  \bibfield{author}{%
  \bibinfo {author} {\bibfnamefont{A.}~\bibnamefont{Sedra}}\ and\ \bibinfo
  {author} {\bibfnamefont{K.}~\bibnamefont{Smith}},\ }%
  \emph{\bibinfo {title} {{Microelectronic Circuits}}},\ \bibinfo {edition}
  {6th}\ ed.\ (\bibinfo {publisher} {Oxford University Press},\ \bibinfo {year}
  {2009})%
  \bibAnnoteFile{NoStop}{NortTheorem}%
\bibitem{Dekker1975}%
  \BibitemOpen
  \bibfield{author}{%
  \bibinfo {author} {\bibfnamefont{H.}~\bibnamefont{Dekker}},\ }%
  \bibfield{journal}{%
  \bibinfo {journal} {Z. Physik B}\ }%
  \textbf{\bibinfo {volume} {21}},\ \bibinfo {pages} {295} (\bibinfo {year}
  {1975})%
  \bibAnnoteFile{NoStop}{Dekker1975}%
\bibitem{Caldeira1983}%
  \BibitemOpen
  \bibfield{author}{%
  \bibinfo {author} {\bibfnamefont{A.}~\bibnamefont{Caldeira}}\ and\ \bibinfo
  {author} {\bibfnamefont{A.}~\bibnamefont{Leggett}},\ }%
  \bibfield{journal}{%
  \bibinfo {journal} {Annals of Physics}\ }%
  \textbf{\bibinfo {volume} {149}},\ \bibinfo {pages} {374} (\bibinfo {year}
  {1983})%
  \bibAnnoteFile{NoStop}{Caldeira1983}%
\bibitem{Johansson06a}%
  \BibitemOpen
  \bibfield{author}{%
  \bibinfo {author} {\bibfnamefont{J.}~\bibnamefont{Johansson}}, \bibinfo
  {author} {\bibfnamefont{S.}~\bibnamefont{Saito}}, \bibinfo {author}
  {\bibfnamefont{T.}~\bibnamefont{Meno}}, \bibinfo {author}
  {\bibfnamefont{H.}~\bibnamefont{Nakano}}, \bibinfo {author}
  {\bibfnamefont{M.}~\bibnamefont{Ueda}}, \bibinfo {author}
  {\bibfnamefont{K.}~\bibnamefont{Semba}},\ and\ \bibinfo {author}
  {\bibfnamefont{H.}~\bibnamefont{Takayanagi}},\ }%
  \bibfield{journal}{%
  \bibinfo {journal} {Phys. Rev. Lett.}\ }%
  \textbf{\bibinfo {volume} {96}},\ \bibinfo {pages} {127006} (\bibinfo {year}
  {2006})%
  \bibAnnoteFile{NoStop}{Johansson06a}%
\bibitem{zhang98a}%
  \BibitemOpen
  \bibfield{author}{%
  \bibinfo {author} {\bibfnamefont{Z.-M.}\ \bibnamefont{Zhang}}, \bibinfo
  {author} {\bibfnamefont{L.-S.}\ \bibnamefont{He}},\ and\ \bibinfo {author}
  {\bibfnamefont{S.-K.}\ \bibnamefont{Zhou}},\ }%
  \bibfield{journal}{%
  \bibinfo {journal} {Phys. Lett. A}\ }%
  \textbf{\bibinfo {volume} {244}},\ \bibinfo {pages} {196} (\bibinfo {year}
  {1998})%
  \bibAnnoteFile{NoStop}{zhang98a}%
\bibitem{pershin10a}%
  \BibitemOpen
  \bibfield{author}{%
  \bibinfo {author} {\bibfnamefont{J.}~\bibnamefont{Martinez-Rincon}}\ and\
  \bibinfo {author} {\bibfnamefont{Y.~V.}\ \bibnamefont{Pershin}},\ }%
  \bibfield{journal}{%
  \bibinfo {journal} {IEEE Trans. Electron. Devices}\ }%
  \textbf{\bibinfo {volume} {58}},\ \bibinfo {pages} {1809} (\bibinfo {year}
  {2011})%
  \bibAnnoteFile{NoStop}{pershin10a}%
\bibitem{rfmemsbook}%
  \BibitemOpen
  \bibfield{author}{%
  \bibinfo {author} {\bibfnamefont{G.~M.}\ \bibnamefont{Rebeiz}},\ }%
  \emph{\bibinfo {title} {{RF MEMS}: Theory, Design, and Technology}},\
  \bibinfo {edition} {1st}\ ed.\ (\bibinfo {publisher} {Wiley-Interscience},\
  \bibinfo {year} {2002})%
  \bibAnnoteFile{NoStop}{rfmemsbook}%
\bibitem{rfmemsbook1}%
  \BibitemOpen
  \bibfield{author}{%
  \bibinfo {author} {\bibfnamefont{V.~K.}\ \bibnamefont{Varadan}}, \bibinfo
  {author} {\bibfnamefont{K.~J.}\ \bibnamefont{Vinoy}}, \bibinfo {author}
  {\bibfnamefont{K.~A.}\ \bibnamefont{Jose}},\ and\ \bibinfo {author}
  {\bibfnamefont{U.}~\bibnamefont{Zoelzer}},\ }%
  \emph{\bibinfo {title} {{RF MEMS} and their applications}},\ \bibinfo
  {edition} {1st}\ ed.\ (\bibinfo {publisher} {Wiley},\ \bibinfo {year}
  {2002})%
  \bibAnnoteFile{NoStop}{rfmemsbook1}%
\bibitem{evoy04a}%
  \BibitemOpen
  \bibfield{author}{%
  \bibinfo {author} {\bibfnamefont{S.}~\bibnamefont{Evoy}}, \bibinfo {author}
  {\bibfnamefont{M.}~\bibnamefont{Duemling}},\ and\ \bibinfo {author}
  {\bibfnamefont{T.}~\bibnamefont{Jaruhar}},\ }%
  in\ \emph{\bibinfo {booktitle} {Introduction to Nanoscale Science and
  Technology}},\ \bibinfo {editor} {edited by\ \bibinfo {editor}
  {\bibfnamefont{M.}~\bibnamefont{Di~Ventra}}, \bibinfo {editor}
  {\bibfnamefont{S.}~\bibnamefont{Evoy}},\ and\ \bibinfo {editor}
  {\bibfnamefont{J.~R.}\ \bibnamefont{Heflin}}}\ (\bibinfo {publisher}
  {Springer},\ \bibinfo {year} {2004})\ pp.\ \bibinfo {pages} {389--416}%
  \bibAnnoteFile{NoStop}{evoy04a}%
\bibitem{mandel95a}%
  \BibitemOpen
  \bibfield{author}{%
  \bibinfo {author} {\bibfnamefont{L.}~\bibnamefont{Mandel}}\ and\ \bibinfo
  {author} {\bibfnamefont{E.}~\bibnamefont{Wolf}},\ }%
  \emph{\bibinfo {title} {Optical Coherence and Quantum Optics}}\ (\bibinfo
  {publisher} {Cambridge University Press},\ \bibinfo {year} {1995})%
  \bibAnnoteFile{NoStop}{mandel95a}%
\bibitem{Watanabe03a}%
  \BibitemOpen
  \bibfield{author}{%
  \bibinfo {author} {\bibfnamefont{M.}~\bibnamefont{Watanabe}}\ and\ \bibinfo
  {author} {\bibfnamefont{D.~B.}\ \bibnamefont{Haviland}},\ }%
  \bibfield{journal}{%
  \bibinfo {journal} {Phys. Rev. B}\ }%
  \textbf{\bibinfo {volume} {67}},\ \bibinfo {pages} {094505} (\bibinfo {year}
  {2003})%
  \bibAnnoteFile{NoStop}{Watanabe03a}%
\bibitem{SQUID_book}%
  \BibitemOpen
  \bibfield{author}{%
  \bibinfo {author} {\bibfnamefont{J.}~\bibnamefont{Clarke}}\ and\ \bibinfo
  {author} {\bibfnamefont{A.}~\bibnamefont{Braginski}},\ }%
  \emph{\bibinfo {title} {{The SQUID Handbook, Volume 2: Applications of SQUIDs
  and SQUID Systems}}},\ \bibinfo {edition} {1st}\ ed.\ (\bibinfo {publisher}
  {Wiley-VCH},\ \bibinfo {year} {2006})%
  \bibAnnoteFile{NoStop}{SQUID_book}%
\bibitem{partensky2002-1}%
  \BibitemOpen
  \bibfield{author}{%
  \bibinfo {author} {\bibfnamefont{M.~B.}\ \bibnamefont{Partensky}},\ }%
  \bibfield{journal}{%
  \bibinfo {journal} {arXiv:physics/0208048}}%
   (\bibinfo {year} {2002})%
  \bibAnnoteFile{NoStop}{partensky2002-1}%
\bibitem{Path_int_book}%
  \BibitemOpen
  \bibfield{author}{%
  \bibinfo {author} {\bibfnamefont{A.}~\bibnamefont{Altland}}\ and\ \bibinfo
  {author} {\bibfnamefont{B.}~\bibnamefont{Simons}},\ }%
  \emph{\bibinfo {title} {Condensed Matter Field Theory}},\ \bibinfo {edition}
  {1st}\ ed.\ (\bibinfo {publisher} {Cambridge University Press},\ \bibinfo
  {year} {2010})%
  \bibAnnoteFile{NoStop}{Path_int_book}%
\end{thebibliography}%
\end{document}